\documentclass{aa}  
\usepackage{caption}
\usepackage{subcaption}
\usepackage[flushleft]{threeparttable}
\usepackage{graphicx}
\usepackage{txfonts}
\usepackage{xcolor}
\usepackage{euscript}

\begin{document} 
 \title{LOFAR Deep Fields: Probing the sub-mJy regime of polarized extragalactic sources in ELAIS-N1}
\subtitle{I. The catalog}
   \author{S. Piras\inst{\ref{inst1}}
         \and C. Horellou\inst{\ref{inst2}}
         \and J.~E. Conway\inst{\ref{inst2}}
         \and M. Thomasson\inst{\ref{inst1}}
         \and S. del Palacio\inst{\ref{inst1}}
         \and T.~W. Shimwell\inst{\ref{inst3},\ref{inst4}}
         \and S.~P. O'Sullivan\inst{\ref{inst5}}
         \and E. Carretti\inst{\ref{inst6}}
         \and I. Šnidari\'c\inst{\ref{inst7}}
         \and V. Jeli\'c\inst{\ref{inst7}}   
         \and B. Adebahr\inst{\ref{inst8}} 
         \and A. Berger\inst{\ref{inst8}}
         \and P.~N. Best\inst{\ref{inst11}} 
         \and M. Br\"uggen\inst{\ref{inst9}}
         \and N. Herrera Ruiz 
         \and R. Paladino\inst{\ref{inst6}} 
         \and I. Prandoni\inst{\ref{inst6}} 
         \and J. Sabater 
         \and V. Vacca\inst{\ref{inst10}}           
          }

    \institute{Department of Space, Earth and Environment, Chalmers University of Technology, 412 96 Gothenburg, Sweden\label{inst1}
  \and Department of Space, Earth and Environment, Chalmers University of Technology, Onsala Space Observatory, 43992 Onsala, Sweden\label{inst2} 
  \and ASTRON, Netherlands Institute for Radio Astronomy, Oude Hoogeveensedijk 4, 7991 PD, Dwingeloo, The Netherlands\label{inst3}  
  \and Leiden Observatory, Leiden University, P.O. Box 9513, 2300 RA Leiden, The Netherlands\label{inst4}  
  \and Departamento de Física de la Tierra y Astrofísica \& IPARCOS-UCM, Universidad Complutense de Madrid, 28040 Madrid, Spain\label{inst5}
  \and INAF Istituto di Radioastronomia, Via Gobetti 101, I-40129 Bologna, Italy\label{inst6} 
  \and Ru{\dj}er Bo\v{s}kovi\'c Institute, Bijeni\v{c}ka cesta 54, 10000 Zagreb, Croatia\label{inst7} 
  \and Ruhr University Bochum, Faculty of Physics and Astronomy, Astronomical Institute, Universit\"atstrasse 150, 44801 Bochum, Germany\label{inst8}
  \and Institute for Astronomy, University of Edinburgh, Royal Observatory, Blackford Hill, Edinburgh, EH9 3HJ, UK\label{inst11} 
  \and Hamburger Sternwarte, University of Hamburg, Gojenbergsweg 112, 21029 Hamburg, Germany\label{inst9}
  \and INAF-Osservatorio Astronomico di Cagliari, Via della Scienza 5, I-09047 Selargius (CA), Italy\label{inst10}
             }

  \date{Received ...} 
  \abstract
   {Quantifying the number density and physical characteristics of extragalactic polarized sources is important for the successful planning of future studies based on Faraday rotation measure (RM) grids of polarized sources to probe foreground Galactic and intergalactic magnetic fields. 
   However, it is proving very hard to detect polarized signal from the population of very faint (sub-mJy) polarized sources at low radio frequencies, and their properties are mostly unknown.
   LOFAR can play an important role in such studies thanks to its sensitivity and angular resolution, combined with the precision on the inferred RM values that can be achieved through low-frequency broad-band polarimetry.    
   }
   {The aim of this study is to probe the sub-mJy polarized source population with LOFAR. In this first paper, we present the method used to stack LOFAR polarization datasets, the resulting catalog of polarized sources, and the derived polarized source counts.
   }
   {The European Large Area ISO Survey-North 1 (ELAIS-N1) field, one of the deepest of the LOFAR Two-Metre Sky Survey (LoTSS) Deep Fields so far, was selected for a polarimetric study at 114.9–177.4 MHz.
   A total area of 25 deg$^2$ was imaged at 6"-resolution in the Stokes $Q$ and $U$ parameters. 
   Alignment of polarization angles was done both in frequency and in Faraday space before stacking datasets from 19 eight-hour-long epochs taken in two different LOFAR observing cycles. A search for polarized sources was carried out in the final, stacked dataset, and the properties of the detected sources were examined. The depolarization level of sources known to be polarized at 1.4 GHz was quantified.
   }
   {A one-sigma noise level, $\sigma_{\rm QU}$, of 19~$\mu$Jy/beam was reached in the central part of the field after stacking. 
   Twenty-five polarized sources were detected above 8$\sigma_{\rm{QU}}$, five of which had not been detected in polarization at any other radio frequencies before. 
   Seven additional polarized components were found by lowering the threshold to 6$\sigma_{\rm{QU}}$ at positions corresponding to sources known to be polarized at 1.4~GHz. 
   In two radio galaxies, polarization was detected from both radio lobes, so the final number of associated radio continuum sources is 31. 
   The detected sources are weakly polarized, with a median degree of polarization of 1.75\% for the sample of sources detected in polarized emission. 
For the 10 polarized sources previously identified in a pilot LOFAR study of the ELAIS-N1 field at 20"-resolution, the RM values are consistent but the degrees of polarization are higher in the 6"-resolution data. The sources previously detected in polarization at 1.4 GHz are significantly depolarized at 150 MHz. The catalog is used to derive the polarized source counts at 150~MHz. 
}
   {This is the deepest and highest-resolution polarization study at 150 MHz to date. A full characterization of the sources and an analysis of the catalog will be presented in Paper II. 
   }  


   \maketitle

\section{Introduction}
\label{sec:int}

The measurement of polarized radio emission from extragalactic sources provides information not only on the polarization properties of the sources themselves but also on the properties of the intervening medium, through Faraday rotation effects, which are related to the distribution of thermal electrons and magnetic fields along the line of sight  
(which goes through any intervening intergalactic medium (IGM), galactic and interstellar medium, including that of the Milky Way). 
Faraday rotation causes the polarization angle $\chi$ of the linearly polarized wave emitted by a source to rotate as it propagates through a magneto-ionic medium:
\begin{equation}
\chi = \chi_0 + {\rm RM}\, \lambda^2 \, ,
\end{equation}
where $\chi$ is measured at the wavelength $\lambda$ of the observation, 
$\chi_0$ is the intrinsic value of the polarization angle, 
and RM is the rotation measure.
In the simplest possible scenario where Faraday rotation occurs in a purely thermal foreground medium, RM is equal to a physical quantity, $\phi(L)$, the Faraday depth of the source at distance $L$ from the observer: 
\begin{equation}
\left( \frac{\phi(L)}{{\rm rad~m}^{-2}} \right) 
= 0.812 \int_{\ell = 0}^{\ell = L} 
\left( \frac{n_{\rm e} (\ell)} { {\rm cm}^{-3} }\right) 
\left( \frac{B_{\parallel}(\ell)} {\mu{\rm G}} \right)
\left( \frac{{\rm d}\ell}{\rm pc}  \right) 
\, , 
\label{eqphi}
\end{equation}
where $n_{\rm e}$ is the density of thermal electrons, 
$B_{\parallel}$ is the line-of-sight component of the magnetic field, $\vec{B}$, taken to be positive if $\vec{B}$ points from the source toward the observer, 
and ${\rm d}\ell$ is the infinitesimal pathlength along the line of sight from the source at $\ell = 0$ to the observer at distance $L$ from the source (e.g., \citealt{Burn1966}; \citealt{ferriere2021}). 

The detection of polarization from extragalactic sources across the sky makes it possible to obtain information on the  properties of the IGM through the construction of so-called “RM grids”. The denser the grid, the finer the reconstruction of IGM polarization structures.
The largest RM catalog available so far covers the entire sky north of $-$40$^\circ$ declination at 1.4~GHz \citep{taylor2009} and contains an average of one polarized source per square degree. It was based on the NRAO Very Large Array Sky Survey 
(NVSS; \citealt{condon1998}). Stacking polarized intensities from NVSS sources, \cite{Stil2014} found a gradual increase in median fractional polarization toward fainter sources. Earlier studies of bright steep-spectrum NVSS sources had shown that the median fractional polarization was higher in low flux-density bins than in high flux-density bins (\citealt{Mesa2002}, \citealt{Tucci2004}).
The behavior of fractional polarization vs flux density at low flux densities is still an open question and dictates the number of polarized sources that will be detected in future polarization surveys, such as with the Square Kilometre Array (SKA). 
\cite{2023ApJS..267...28V} 
recently published an RM catalog of 55\,819 sources (including Galactic sources) gathered from 42 published catalogs, and proposed standards for reporting polarization and RM measurements. 

Deeper polarization surveys will produce denser RM grids. 
The deepest extragalactic polarization studies at 1.4~GHz of the northern sky are those of ELAIS-N1\footnote{
the European Large-Area ISO Survey-North 1 field} 
(\citealt{grant2010}),
of the Lockman Hole (\citealt{Berger2021}), 
and of GOODS-N (\citealt{Rudnick2014}). 
Deep polarization surveys of the southern sky, also at 1.4~GHz, were carried out in ELAIS-S1 and CDF-S \citep{Hales2014_440}. 
Those surveys vary greatly in sky coverage (about 15 deg$^2$ for ELAIS-N1, 6.5~deg$^2$ for the Lockman Hole, and 0.3~deg$^2$ for GOODS-N) and in depth, the deepest survey being that of the Lockman hole with a one-sigma sensitivity of 7~$\mu$Jy beam$^{-1}$, and a beam of 15$\arcsec$ \citep{Berger2021}. 
The largest number density of polarized sources (51 per square degree) was found in the small GOODS-N field that was observed at higher angular resolutions 
(1.6$\arcsec$ and 10$\arcsec$; \citealt{Rudnick2014}). 

The aim of the LOFAR Magnetism Key Science Project (MKSP)\footnote{\url{http://lofar-mksp.org}{http://lofar-mksp.org}} is to investigate the magnetized Universe using LOFAR. 
The search for polarized sources at low radio frequencies is particularly arduous due to the influence of the ionosphere (e.g., \citealt {Sotomayor2013})
and Faraday depolarization effects which become increasingly strong towards lower frequencies 
because of the large amounts of Faraday rotation suffered by different regions of a synchrotron-emitting source within the telescope beam (e.g. \citealt{Burn1966, Sokoloff1998}). 
Some of these difficulties can, however, be overcome with LOFAR thanks to the angular resolution that reduces beam depolarization, the high precision on rotation measures (approximately 1~rad~m$^{-2}$) that can be achieved through broadband polarimetry, and the sensitivity. 

Polarization studies with LOFAR at 150~MHz in the field of nearby galaxy M51 resulted in the detection of 
0.3~polarized sources per square degree at $20 \arcsec$ resolution and one-sigma sensitivity of 100~$\mu$Jy beam$^{-1}$ 
(\citealt{mulcahy2014}, \citealt{Neld2018}). 
By analyzing the 570~deg$^2$ region presented in the Preliminary Data Release from the LOFAR Two-Metre Sky Survey (LoTSS, \citealt{Shimwell2017, Shimwell2019}) \cite{Vaneck2019} found 92~polarized radio sources in images of $4.3'$-resolution and 1~mJy~beam$^{-1}$ one-sigma sensitivity. 
Polarization was detected in a number of giant radio galaxies 
(\citealt{OSullivan2018}, \citealt{OSullivan2019}, 
\citealt{Stuardi2020}, \citealt{Mahatma2021}), an indication that radio sources located in low-density environments have a greater chance to survive depolarization at low frequencies. 
\cite{OSullivan2020} used the RM differences measured between close physical and non-physical pairs of radio sources in LoTSS data to place constraints on the magnetization of the cosmic web. 
A catalog of 2461 polarized sources with RM values across 5720~deg$^2$ of the northern sky was produced by \cite{lotssdr2} who used the second data release (DR2) of the LoTSS at 20$''$~resolution. \cite{Carretti2022,Carretti2023} used this catalog to probe the strength and evolution of magnetic fields in cosmic filaments.

The LOFAR Deep Fields (Bo\"otes, the Lockman Hole and ELAIS-N1; \citealt{Tasse2021}, \citealt{Sabater2021}), 
along with deep observations of GOODS-N (Vacca et al., in prep), are best suited for deep polarization searches.  
To date, the deepest published polarization study with LOFAR was carried out on the ELAIS-N1 field by \cite{HerreraRuiz2021}, who developed a method to stack polarization data taken at different epochs. 
By combining six datasets, each from eight-hour-long observations, they were able to reach a one-sigma sensitivity of 26~$\mu$Jy beam$^{-1}$ in the central part of the final image of the field at a resolution of $20\arcsec$, which enabled the detection of 10 polarized sources in an area of 16~deg$^2$ (0.6 polarized source per square degree). 

In this work, we expand on the pilot study of \cite{HerreraRuiz2021} by improving on both the angular resolution and the sensitivity, and enlarging the analyzed field area; this was achieved by re-imaging the data at a resolution of 6 arcseconds, stacking data from 19 different epochs, and imaging a field of 25~deg$^2$.  
The paper is organized as follows. 
After summarizing the existing polarization studies of the ELAIS-N1 field in Sect.~\ref{sect2}, we present 
in Sect.~\ref{sec:data} the observations, the imaging and calibration procedures, the datasets, and the data processing to produce Faraday cubes. 
The stacking method is described in Sect.~\ref{sec:stacking}
and the search for polarized sources in the stacked data in Sect.~\ref{sec:source_finding}. 
In Sect.~\ref{sec:results} we present and discuss the results, and conclude 
in Sect. \ref{sec:conclusion}. 
Further analysis of the catalog of polarized sources will be presented in a companion paper (Piras et al., Paper II).

\section{Polarization searches in the ELAIS-N1 field}
\label{sect2}

\begin{figure*}
\includegraphics[width=\linewidth]{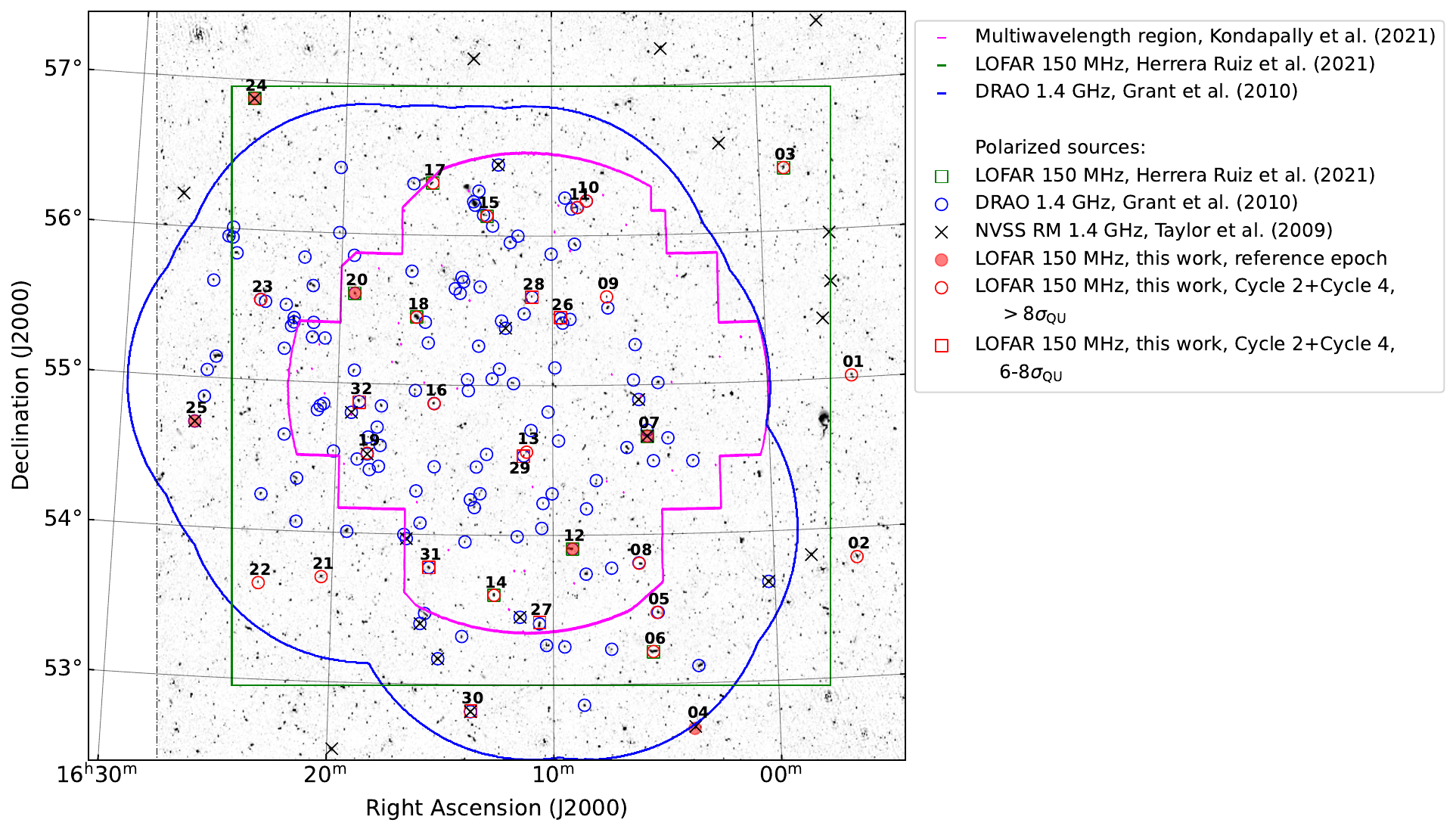}
\caption{LOFAR 150~MHz total-intensity image of the ELAIS-N1 field at 6$''$-resolution and of size 5$\times$5~deg$^2$ \citep{Sabater2021}, on which the coverages of various polarization surveys and the locations of polarized sources are overlaid, as indicated in the legend.} 

\label{fig:field} 
\end{figure*}

\begin{table*}[ht]
\caption{Basic characteristics of polarization studies covering the ELAIS-N1 field.}  
\label{table:PolSourcesSurveys}   
\centering                         
\begin{tabular}{c c c c c c c}   
\hline\hline                
 Catalog & Reference & Frequency & Sensitivity in $Q$ and $U$ & Resolution     & Area  & Number of polarized sources\\ 
         &      &           &($\mu$Jy beam$^{-1}$)   & (arcsec)       & (deg$^2$)\\
\hline
NVSS RM             &(1)    & 1.4 GHz & 290 &  45 & 25 & 25\\
DRAO ELAIS-N1       &(2)    & 1.4 GHz & 78 & 49$\times$59 &7.43 & 83\\
DRAO ELAIS-N1       &(3)    & 1.4 GHz & 45 & 49$\times$62 & 15.16 & 136\\
LOFAR               &(4)    & 150 MHz & 26 & 20 & 16  & 10\\
LOFAR               &(5)    & 150 MHz & 22 & 6  & 25  & 31\\
\hline                                   
\end{tabular}
\tablefoot{
(1) RM catalog from \cite{taylor2009} based on the NVSS \citep{condon1998}; 
number of RM entries within the 25-square-degree area of ELAIS-N1; 
(2) \cite{taylor2007}; 
(3) \cite{grant2010}; the resolution varies with declination across the mosaic as 49$''\times49''/\sin\delta$;  the listed value is the resolution at the center of the mosaic;  
(4) \cite{HerreraRuiz2021};
(5) This work. 
}
\end{table*}
The ELAIS-N1 field is a region of the northern hemisphere\footnote{
RA(J2000) = $16^{\rm h}10^{\rm m}01^{\rm s}$, Dec(J2000) = 54$^\circ 30' 36''$; 
Galactic longitude and latitude $(l,b) = (84^\circ, 45^\circ$)} 
observed by the ELAIS survey \citep{oliver2000}, originally chosen for deep extragalactic observations with the Infrared Space Observatory (ISO) because of its low infrared background \citep{Rowan-Robinson2004, vaccari2005}. 
The ELAIS-N1 field has been the target of a number of polarization studies.
The Galactic polarized foregrounds were imaged with LOFAR by 
\cite{Jelic2014}, and more recently by \cite{Snidaric2023} who used 150~hours of LOFAR observations at very low resolution (4.3~arcmin) to study the diffuse Galactic polarized emission. 
Extragalactic polarized sources were detected in different surveys, as summarized in Table~\ref{table:PolSourcesSurveys}. 
Deep polarization imaging has been carried out at 1.4~GHz by \cite{taylor2007} and \cite{grant2010} at the Dominion Radio Astronomical Observatory (DRAO).
Of particular interest are the results from \cite{grant2010} that can be used to compare the characteristics of the polarized sources at 1.4~GHz and at 150~MHz and investigate depolarization between those two frequencies in the region of overlap (about 15 square degrees).
\cite{grant2010} detected 136 polarized sources and used the catalog to construct the Euclidean-normalized polarized differential source counts down to 400~$\mu$Jy. 
They found, at 1.4~GHz, that fainter radio sources have a higher fractional polarization than the brighter ones. 
This catalog, however, does not contain any information on rotation measures, and for this we rely on the RM catalog of \cite{taylor2009} derived from data from the NRAO Very Large Array Sky Survey (NVSS; \citealt{condon1998}) taken in two frequency bands centered at 1.365 and 1.435~GHz. 

Figure~\ref{fig:field} shows the central 25~deg$^2$ area of the 6$''$-resolution LOFAR image of the ELAIS-N1 field \citep{Sabater2021}; highlighted in magenta is the central region of the field with ample multiwavelength coverage (see \citealt {kondapally2021} for a summary). 
Also overlaid on Fig.~\ref{fig:field} are the two regions within the field where dedicated polarization searches were performed: at 1.4~GHz with DRAO \citep{grant2010} and at 150~MHz with LOFAR \citep{HerreraRuiz2021}. 
Also shown are the locations of the polarized sources detected in those surveys, of the sources with RM values from the NVSS \citep{taylor2009}, 
and the polarized sources identified in this work.  

\section{Data}
\label{sec:data}
\begin{table*}[t]
\caption{LOFAR observing configurations for the ELAIS-N1 Deep Field and parameters of RM synthesis.}        
\label{table:observations_setup}      
\centering                          
\begin{tabular}{c c c c c c c }        
\hline\hline                
 LOFAR Cycle & Frequency Range  & Number of Frequency Channels & Channel Width & $\delta \phi$  & max-scale & $|\phi_{\rm max}|$\\ 
      & (MHz)  &  & (kHz)  & (rad m$^{-2}$) & (rad m$^{-2}$) & (rad m$^{-2}$)\\
\hline
 2  & 114.979 -- 177.401 & 800 & 78.12500 & 0.9 & 1.1 & 385\\  
 4  & 114.989 -- 177.391 & 640 & 97.65625 & 0.9 & 1.1 & 308\\
\hline                                   
\end{tabular}
\end{table*}
\begin{table*}[ht]
\caption{LOFAR ELAIS-N1 datasets used in this paper.}  
\label{table:observations}      
\centering                          
\begin{tabular}{ l c c c c | l c c c c }
\hline\hline                
 & & Cycle 2 & & &  & & Cycle 4\\
 \hline
ID & LOFAR ID &  Date & Duration & $\sigma_{\rm QU}$ & ID & LOFAR ID  & Date & Duration & $\sigma_{\rm QU}$\\ 
   & & & & $(\mu$Jy beam$^{-1})$ &  & & & & $(\mu$Jy beam$^{-1})$ \\
\hline
009 & L229064 & 2014-05-19 & 8h00m05s & 88 & 020$\ast$ & L345624   & 2015-06-07 & 7h40m06s & 82\\  
010 & L229312 & 2014-05-20 & 8h00m05s & 88 & 021       & L346136   & 2015-06-14 & 7h40m06s & 82\\  
012 & L229673 & 2014-05-26 & 8h00m05s & 77 & 022       & L346154   & 2015-06-12 & 7h40m06s & 96\\
013 & L230461 & 2014-06-02 & 8h00m05s & 109& 023       & L346454   & 2015-06-17 & 7h40m06s & 96\\ 
\textbf{014} & \textbf{L230779}  
              & 2014-06-03 & 8h00m05s & 73 & 024$\ast$ & L347030   & 2015-06-19 & 7h40m06s & 81\\  
015 & L231211 & 2014-06-05 & 8h00m05s & 79 & 026       & L347494   & 2015-06-26 & 7h40m06s & 90\\ 
016 & L231505 & 2014-06-10 & 7h20m06s & 78 & 028$\ast$ & L348512   & 2015-07-01 & 7h40m06s & 104\\
017 & L231647 & 2014-06-12 & 6h59m58s & 84 & 031$\ast$ & L369530   & 2015-08-22 & 7h40m06s & 84\\ 
018 & L232981 & 2014-06-27 & 4h59m58s & 95 & 032       & L369548   & 2015-08-21 & 7h40m06s & 88\\
019 & L233804 & 2014-07-06 & 5h00m01s & 122\\ 
\hline                                   
\end{tabular}
\tablefoot{
Adapted from Table 1 of \cite{Sabater2021}.\\
Columns~1 and 6: internal ID code; the reference epoch (014) is marked in bold;
asterisks indicate the four datasets (out of six) used in the pilot study of \cite{HerreraRuiz2021};\\
Cols~2 and 7: standard LOFAR ID;\\
Cols~3 and 8: date on which the observation started;\\
Cols~4 and 9: total duration of the observation;\\
Cols~5 and 10: noise level in polarization in the 2.5~$\times$~2.5~arcmin$^2$ central region of the image.
}
\end{table*}

\subsection{Observations}
A detailed description of the observations has been given by \cite{Sabater2021}. 
The datasets used in this work\footnote{New observations were analyzed recently, as described fully in \cite{Best2023}, and will be presented in Shimwell et al., in prep.} 
consist of observations of the ELAIS-N1 field taken in full polarization with the LOFAR High Band Antenna (HBA) between May 2013 and 27 August 2015 (Cycles 0, 2, and 4; proposals LC0\_019, LC2\_024, and LC4\_008). 
Defining as epoch an eight-hour LOFAR observation, 22 epochs from Cycles~2 and~4 are available for a total of 176 hours of observations. The observations are centered on RA\,=\,16$^{\rm h}$11$^{\rm m}$00$^{\rm s}$, Dec\,=\,55$^{\circ}$00$^{\prime}$00$^{\prime\prime}$ (J2000). 

As shown in Table~\ref{table:observations_setup}, the frequency setup of the Cycle~4 observations was different from that of Cycle~2. This meant that the stacking of data from different cycles could not be done directly in frequency space; it was done in Faraday space, as will be described in Sect.~\ref{sec:stacking}. 
\cite{Snidaric2023} also stacked data in Faraday space to investigate the Galactic polarized emission in ELAIS-N1, but used a different approach, cross-correlating data from different epochs to determine the offsets. 
The last three parameters in Table~\ref{table:observations_setup} will be defined in Sect.~\ref{sec:RMsyth}. 

\subsection{Data imaging and calibration}

The data for ELAIS-N1 were calibrated for the \cite{Sabater2021} study and reprocessed for the work presented here. For each frequency channel of the data from the 22 epochs we created primary-beam-corrected Stokes $Q$ and $U$ images at $6''$ angular resolution. The images were created with the software {\tt DDFacet} \citep{Tasse2018}, allowing us to apply the direction independent and dependent (45 directions) calibration solutions derived by \cite{Sabater2021} whilst correcting for the LOFAR beam. As described in more detail in \cite{lotssdr2}, 
the calibration solutions were derived with the assumption that $Q=U=V=0$, which has the effect of suppressing instrumental polarization -- although it can produce spurious polarized sources if there are genuinely bright ($> 10$~mJy~beam$^{-1}$) polarized sources in the field, but this is not an issue for the ELAIS-N1 field. 
The images were not deconvolved as this functionality was not available in {\tt DDFacet} at the time of processing. 
For six of the 22 datasets, some frequency channels were missing after the data processing. 
The reason for this is unknown to us, but we suspect that it has to do with the large data volumes and memory issues. 
For three datasets (L229064, L230461, L346154), we could identify the missing channels as this information was present in the original Measurement Sets and those datasets were included in the analysis. 
The remaining three datasets (L229387, L347512, L366792) were excluded as the missing frequency channels could not be identified and the frequency information could not be recovered.

Because of the lack of polarization calibrators at low frequencies, the inferred polarization angles are not absolute. However, this is not an issue in the work presented here that focuses on measurements of Faraday rotation of the polarization angles across the frequency band. Although the data are corrected for ionospheric Faraday rotation during the eight-hour-long integrations, some variations remain between the various datasets and the polarization angles will need to be aligned before stacking. This will be done using one of our brightest polarized sources as a polarization calibrator, as explained in Sect.~\ref{sec:stacking}.

\subsection{Datasets}

In Table~\ref{table:observations} we summarize the 19 observations and datasets used in this paper. Each dataset consists of Stokes $Q$ and $U$ frequency cubes of the imaged region of 25~deg$^2$ area. 
The size of each Stokes-parameter cube is 12005 pixels $\times$ 12005 pixels $\times$ number of frequency channels (800 for datasets from Cycle~2 and 640 for Cycle~4 data). The pixel size in (RA, Dec) is 1.5$'' \times$1.5$''$. 
The data volume of each Stokes-parameter cube is $\sim$400~GB, resulting in a total data volume of $\sim$15~TB for all epochs. Because of the large size of the input data, and estimating that the output data would be about three times larger, the data processing had to be carried out on a supercomputer (Vera, at the Chalmers Centre for Computational Science and Engineering) and a strategy had to be developed to manage memory issues and optimize the processing times. In particular, 
the Stokes $Q$, $U$ frequency cubes were divided in horizontal strips in ranges of declination (Sect.~\ref{sec:source_finding}).

\subsection{Rotation measure synthesis} 
\label{sec:RMsyth}

The complex linear polarization is:
\begin{equation} \label{eq:complexpol_chi}
    \mathcal{P} = Q + {\rm i} U 
    = P\, \mathrm{e}^{2i \chi} \, ,
\end{equation}
where $P$ is the polarized intensity
and $\chi$ the polarization angle. 
All these quantities ($\mathcal{P}$, $Q$, $U$, $P$, $\chi$) depend on frequency, $\nu$, but it is common to express them as a function of wavelength squared, $\lambda^2$, because of the nature of Faraday rotation. In this section we describe how data are transformed from frequency space to Faraday depth space using the rotation measure synthesis \citep{Burn1966, Brentjens2005}. 

The complex linear polarization can be expressed as the integral over all Faraday depths, $\phi$, of the complex Faraday dispersion function $\mathcal{F}(\phi)$ modulated by the Faraday rotation:
\begin{equation}
    \mathcal{P}(\lambda^2) = \int_{-\infty}^{\infty}
    \mathcal{F}(\phi) \, \mathrm{e}^{2{\mathrm i}\phi \lambda^2} \, 
    \mathrm{d} \phi \, .
\end{equation}

This is a Fourier-transform-like relation that can in principle be inverted to express $\mathcal{F}(\phi)$: 
\begin{equation}
    \mathcal{F}(\phi) = \frac{1}{\pi} \int_{-\infty}^{\infty}
    \mathcal{P}(\lambda^2)\, {\mathrm e}^{-2 {\mathrm i} \phi \lambda^2} \, \mathrm{d}\lambda^2\, .
\end{equation}

In practice, because of the limited number of channels at which the polarized intensity can be measured, the inferred Faraday dispersion function is the convolution of the true Faraday dispersion function $\mathcal{F}(\phi)$ with the so-called RM Spread Function (RMSF):
\begin{equation}
 R(\phi)=\frac{\int_{-\infty}^{\infty}
   W(\lambda^2) \, \mathrm{e}^{-2 \mathrm{i} \phi \lambda^2} \, \mathrm{d}\lambda^2}
 {\int_{-\infty}^{\infty} 
     W(\lambda^2)\, \mathrm{d}\lambda^2} \, ,
\end{equation}
where the sampling function or weight function $W(\lambda^2)$ is nonzero at the measured $\lambda^2$ and zero elsewhere.

Fig.~\ref{fig:rmsf} shows two examples of RMSFs for two LOFAR  observation cycles. 
They are very similar despite the slightly different frequency setups during Cycle~2 and Cycle~4. Even within the same observation cycle, the RMSF may be slightly different as it depends on the locations of flagged frequency channels.

Three key parameters of RM synthesis are 
the resolution in Faraday space, the largest scale in Faraday depth and the maximum Faraday depth to which the technique is sensitive: 
\begin{flalign} \label{eq1}
\delta \phi \approx \frac{2 \sqrt{3}}{\Delta \lambda^2} \\
{\rm max\text{-}scale} \approx \frac{\pi}{\lambda^2_{\rm min}}   \\
|\phi_{\rm max}|  \approx \frac{\sqrt{3}}{\delta \lambda^2}\, ,
\end{flalign}
where $\Delta \lambda^2 = \lambda_{\rm max}^2 - \lambda_{\rm min}^2$ is the $\lambda^2$ coverage, 
$\lambda_{\rm max} = c/\nu_{\rm min}$ is the maximum wavelength, 
$\lambda_{\rm min} = c/\nu_{\rm max}$ the minimum wavelength, 
$\nu_{\rm min}$ and $\nu_{\rm max}$ are the lowest and highest frequencies of the whole bandwidth, 
and $\delta \lambda^2$ is the channel width in $\lambda^2$ space \citep{Brentjens2005}. 
In Table~\ref{table:observations_setup} we list the values of those parameters for the LOFAR datasets from each observing cycle. The main difference between the frequency setups of Cycle~2 and Cycle~4 is the width of the frequency channels, which affects $\phi_{\rm max}$. Because the total frequency coverage was very similar during both cycles, the resolution in $\phi$ space, $\delta\phi$, and the largest scale in $\phi$, max-scale, are basically identical. 

In the simplest scenario, in which the synchrotron radiation is Faraday-rotated by a foreground magneto-ionic medium, the measured RM is equal to the Faraday depth. For simplicity, $\phi$ and RM are used here as synonyms.

We used the Python code \texttt{pyrmsynth\_lite}\footnote{\url{https:/github.com/sabourke/pyrmsynth_lite}} to perform RM synthesis \citep{Brentjens2005}, with uniform weighting and with the {\tt RMCLEAN} option \citep{Heald2009} to deconvolve the Faraday dispersion function from the RMSF. The outputs of {\tt pyrmsynth\_lite} are three-dimensional Faraday cubes (RA, Dec, $\phi$) and two-dimensional sky maps: 

\begin{itemize}
    \item The Stokes $Q$ and $U$ Faraday cubes ($Q(\phi)$, $U(\phi)$) are the real and imaginary parts of the reconstructed Faraday dispersion function;  
    \item The polarized intensity cube ($F(\phi) = \sqrt{Q(\phi)^2 + U(\phi)^2}$);
    \item The corresponding two-dimensional maps, obtained at values of $\phi$ that correspond to the peaks of $F(\phi)$;   
    we name these maps $F_{\rm map}$, $Q_{\rm map}$, $U_{\rm map}$, $RM_{\rm map}$.
\end{itemize}

The Faraday cubes have a span in Faraday depth of $\pm$450~rad~m$^{-2}$ and a spacing of 0.3~rad~m$^{-2}$. 
The range $[-3,+1.5]$~rad~m$^{-2}$
was excluded from the analysis to avoid the leakage signal 
(following \citealt{HerreraRuiz2021}). 
Additionally, noise maps were created with pixel values equal to the mean of the noise levels in  $Q(\phi)$ and $U(\phi)$, 
calculated as the standard deviations in the [350, 450]~rad~m$^{-2}$ range of Faraday depths, where no polarized signal is expected.

\section{Stacking} \label{sec:stacking}

\begin{figure*}
\centering
\includegraphics[height=23cm]{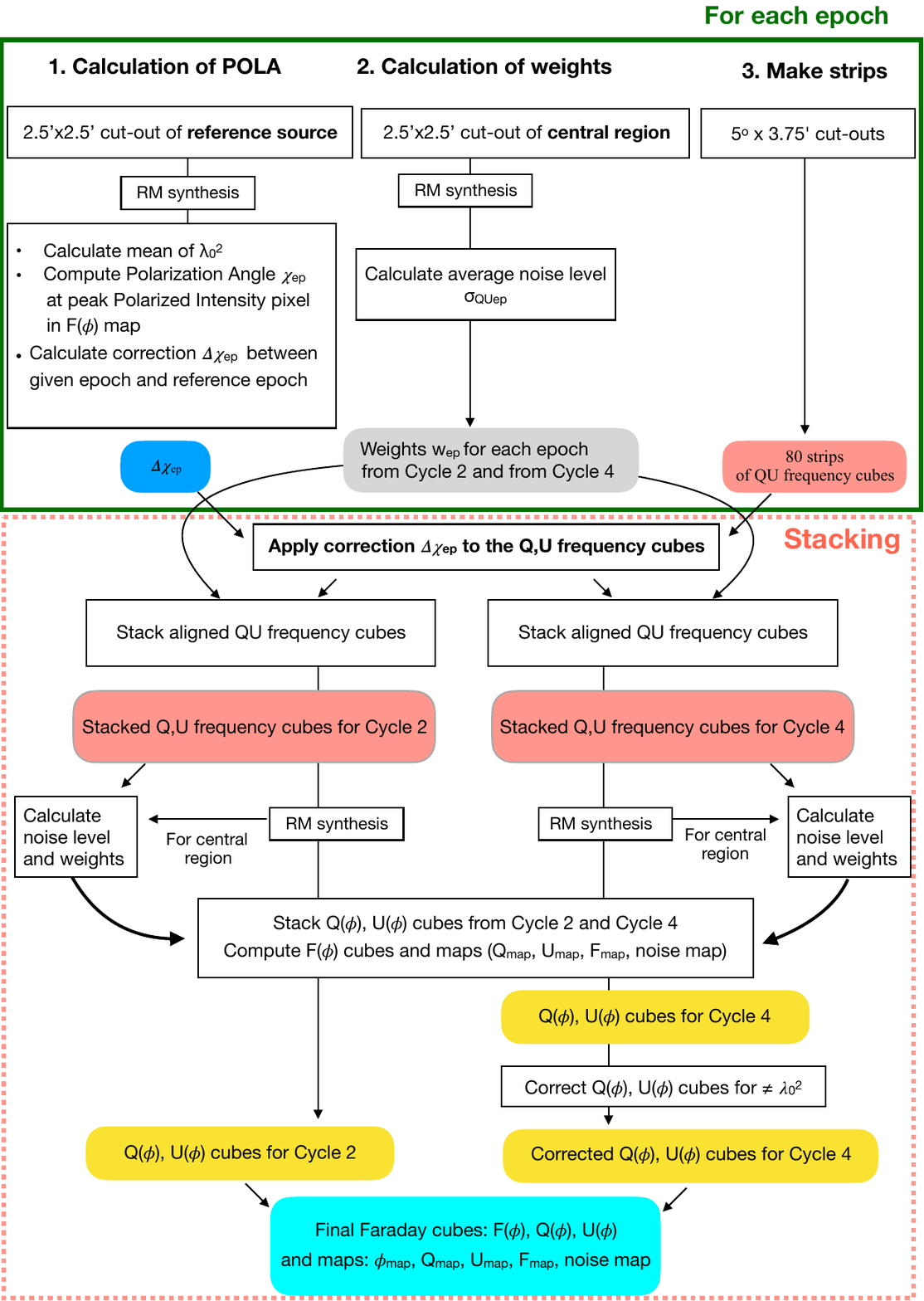}
\caption{Flowchart of the stacking process. POLA stands for Polarization Angle.}
\label{fig:flowchart} 
\end{figure*}

The stacking procedure is summarized in Figure~\ref{fig:flowchart}. 
Although the data from each given epoch were corrected for ionospheric Faraday rotation, 
polarization angles from different observing runs may differ due to ionospheric correction effects, as a full polarization calibration (for the ensemble of 19 datasets) was not performed.  
This can lead to depolarization when data from different epochs are combined. It is therefore crucial to align the polarization angles before stacking.
Because of the different frequency setups in Cycle~2 and Cycle~4, the stacking had to be done in two steps: first, data from a given cycle were stacked in frequency space, following the method used by \cite{HerreraRuiz2021}; then the two stacked datasets from both cycles were stacked in Faraday depth space.

\subsection{Stacking frequency cubes from same observation cycle} \label{sec:stacking-Noelia}

\subsubsection{The reference source: calculating corrections}
\label{sec:refsource}
\begin{figure*}[!ht] 
\centering
\includegraphics[height=6.9cm]{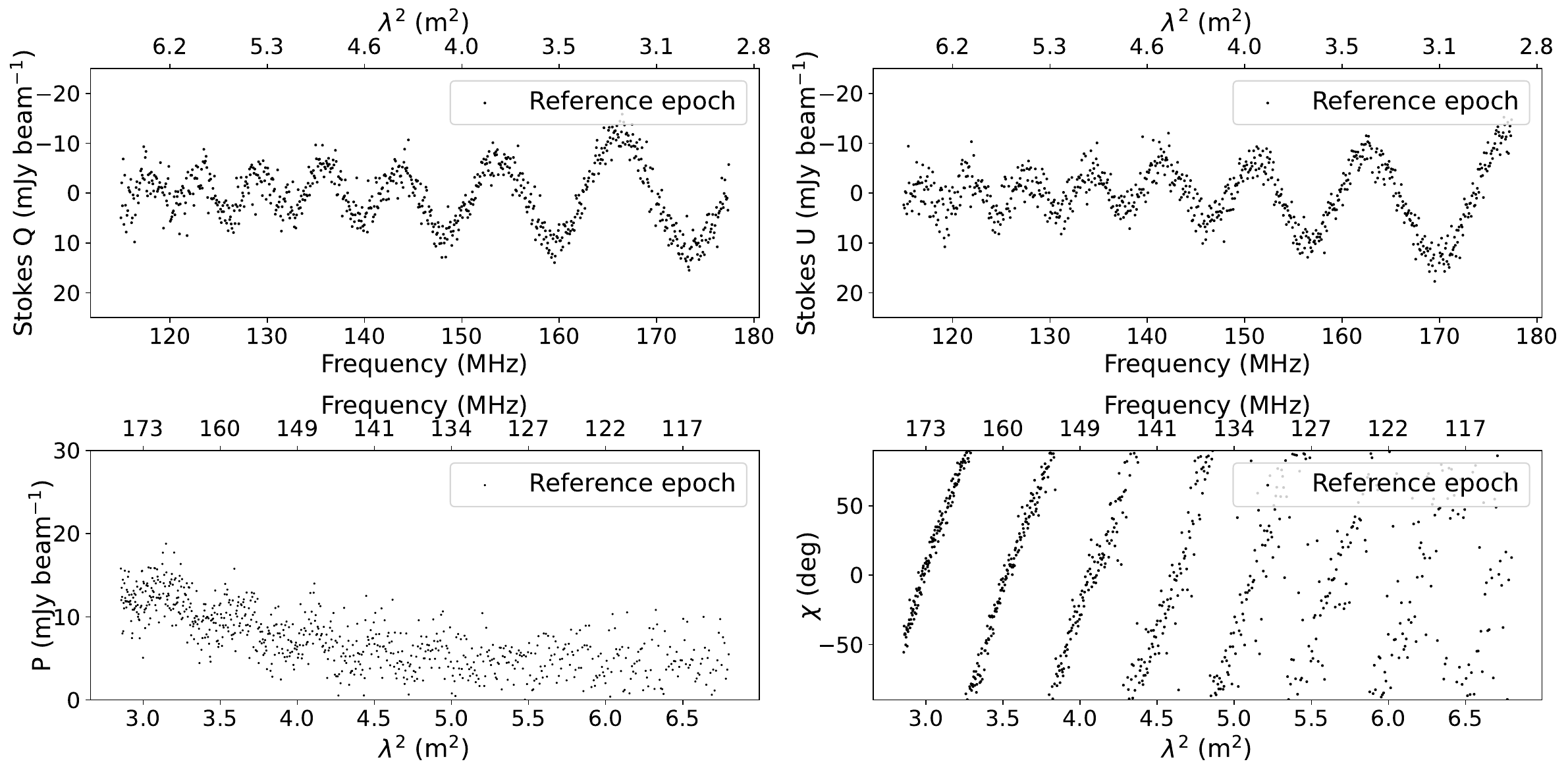} 
\includegraphics[height=6.9cm]{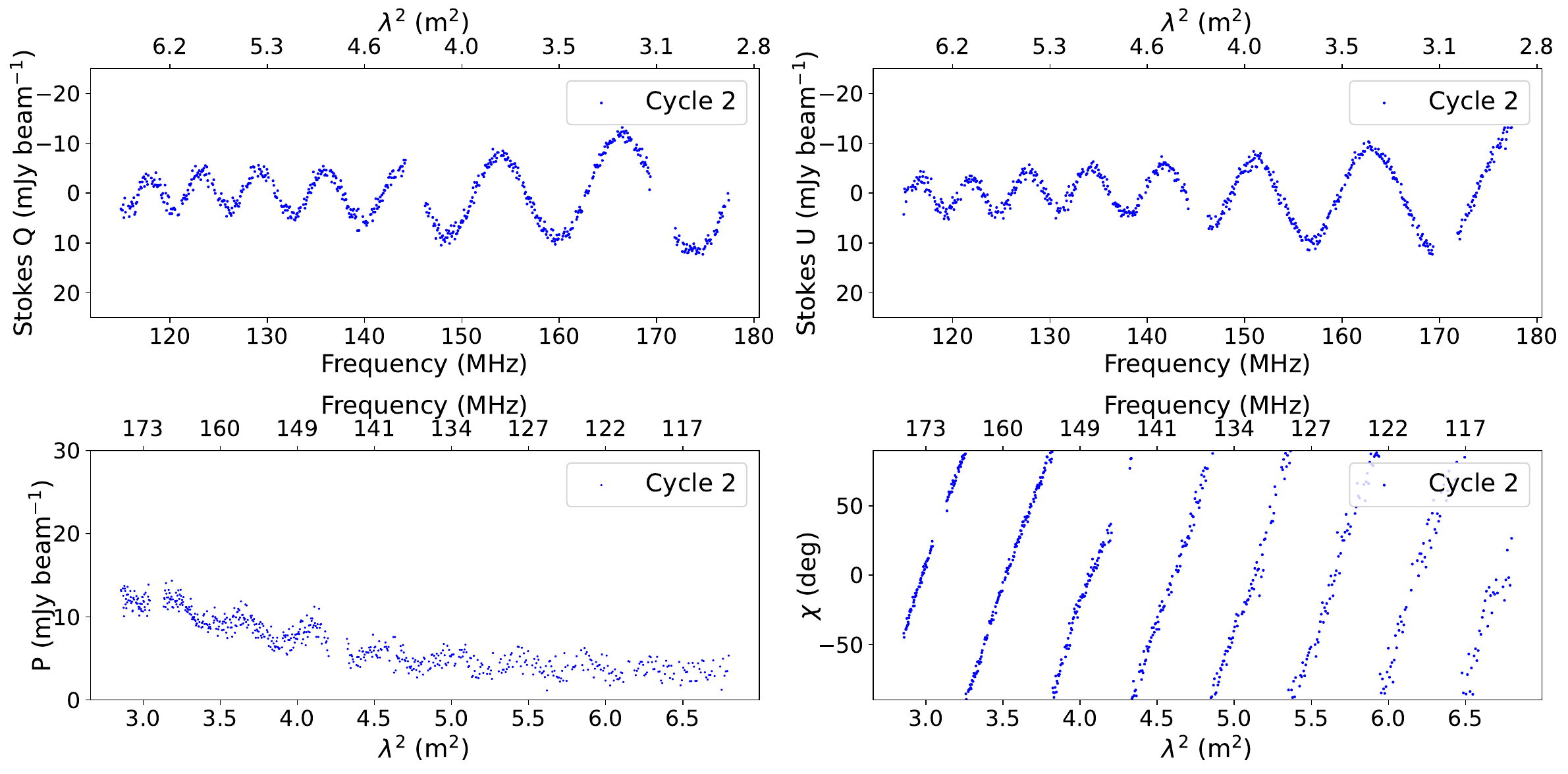}
\includegraphics[height=6.9cm]{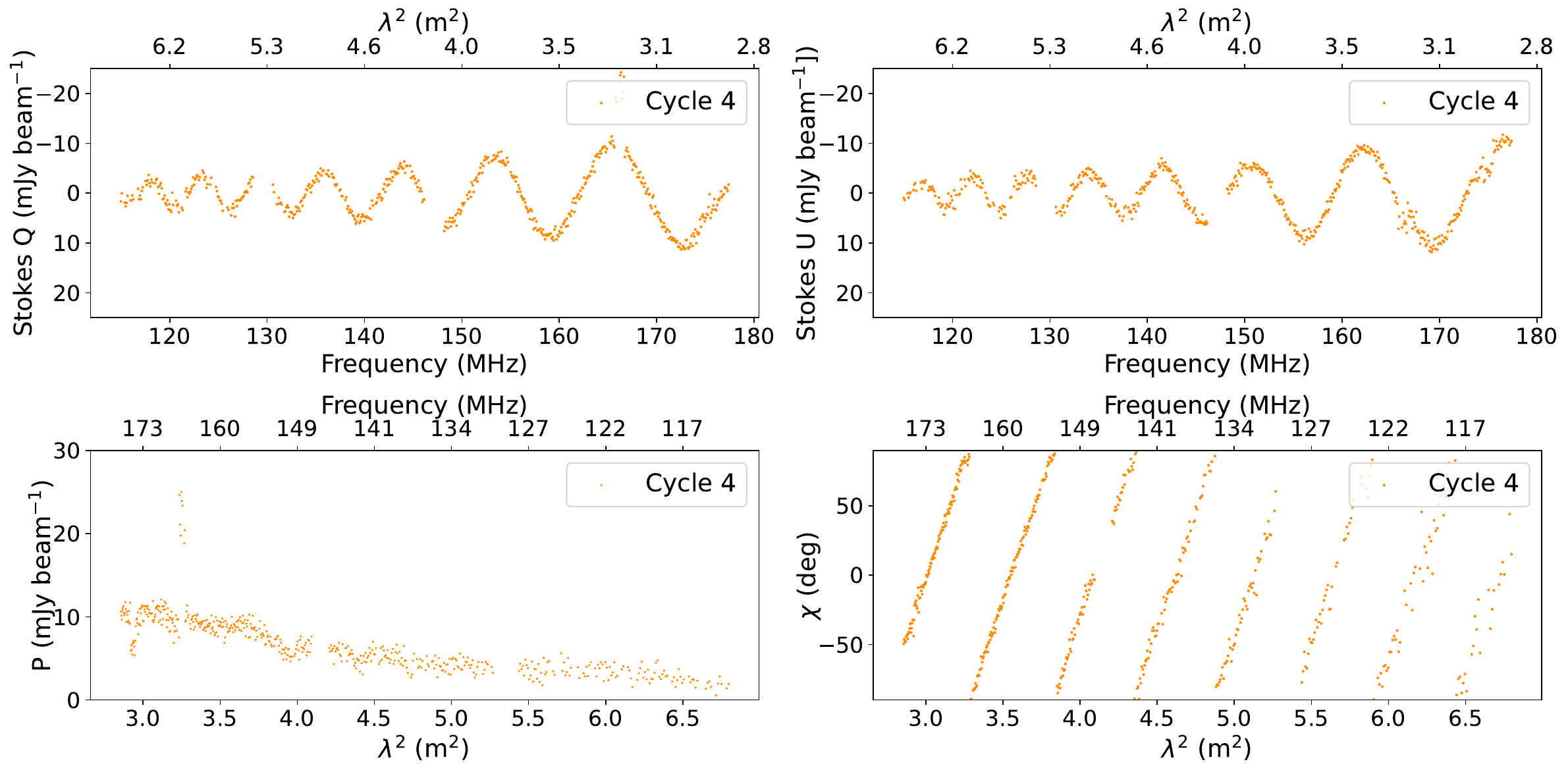}
\caption{Polarization characteristics of the reference source (Source~07). Stokes $Q$ and $U$ intensities, polarized intensity $P$ and polarization angle $\chi$ are shown as a function of frequency and wavelength squared $\lambda^2$ (bottom and top axes) for the reference epoch (in black, first two rows), 
and for the stacked data from Cycle~2, 
(in blue, third and fourth row) 
and from Cycle~4 
(in orange, two last rows). 
The scatter in the data is reduced by stacking.
The effect of Faraday rotation is clearly visible in the variations of $Q$, $U$ and $\chi$; 
the decrease of the polarized intensity with $\lambda^2$ is an indication of Faraday depolarization. 
The final combination of Cycle~2 and Cycle~4 data is done in Faraday depth space, as described in the text.
}
\label{fig:ref_source} 
\end{figure*}

The stacking method is based on the use of a polarized reference source in a reference epoch as polarization angle calibrator. For each epoch, angular corrections are calculated for the reference source and applied to the whole field. 

As {\underline{reference source}} we used the same polarized source as the one selected by \cite{HerreraRuiz2021} in their LOFAR pilot polarization study of ELAIS-N1 (ILTJ160538.33+543922.6; their source 02, our source~07). 
This source has a peak polarized intensity of $\sim$6~mJy~beam$^{-1}$ and an RM of $\sim$6~rad~m$^{-2}$ in data from all epochs. 
It is located in the part of the ELAIS-N1 field for which value-added data are available, including a photometric redshift estimate of 0.7911 \citep{Duncan2021}. 
In the LOFAR 6-arcsec Stokes $I$ image of \cite{Sabater2021}, there are two radio components on either side of the host galaxy, 
and polarization is detected in the south-western component which is probably a radio lobe. As will be shown later, this source is the one in which we detect polarization with the highest signal-to-noise ratio (S/N). It is well detected at each epoch. 

Cut-outs of 2.5~$\times$~2.5~arcmin$^2$ in the data cubes centered at the reference source were made for each epoch. RM synthesis was performed on each cut-out.  
To achieve a better precision on the RM values of the reference source, 
RM synthesis was performed with a sampling of 0.01~rad~m$^{-2}$ around the RM value of the reference source in the range of $\pm$10 rad m$^{-2}$.

As {\underline{reference epoch}} we used Epoch~014 from Cycle~2, which is the dataset of highest quality, with the lowest and most uniform noise. \cite{HerreraRuiz2021} had stacked only data from Cycle~4 and used Epoch~24 as their reference epoch. 

The {\underline{polarization angle calibrator}} is the reference source at the reference epoch. Its coordinates\footnote{
The polarization angle calibrator has J2000 coordinates RA\,=\,241.4080$^\circ$ and Dec\,=\,54.6551$^\circ$}
are those of the pixel of the peak intensity in the polarized intensity map. 

In Fig.~\ref{fig:ref_source} we show the polarization characteristics of the polarization angle calibrator, and the same quantities for the corresponding pixels in stacked data from Cycle~2 and from Cycle~4 for the reference source. The $Q$ and $U$ Stokes parameters show clear oscillations with frequency (or wavelength squared), as expected with Faraday rotation. The polarized intensity decreases with increasing $\lambda^2$, which is a sign of Faraday depolarization. The polarization angle ($\chi$, bottom right, and corresponding panels above) varies linearly with $\lambda^2$. The positive slope indicates a positive RM. These graphs show clearly that stacking reduced the scatter in the data. 

Each dataset has a slightly different frequency coverage, and therefore the mean frequency and the corresponding wavelength squared ($\lambda_0^2$) are slightly different, as listed in Table~\ref{table:shift}. This means that the polarization angles output from RM synthesis refer to a different $\lambda_0^2$ for each epoch. Therefore, the polarization angles must be corrected to take into account the Faraday rotation, RM$\cdot\Delta\lambda_0^2$, that occurs between two different values of $\lambda_0^2$.

\begin{table}[t]
\caption{Values of $\lambda_0^2$ in datasets of different epochs, RM values of the reference source and applied polarization angle corrections. The reference epoch is indicated in bold.
The horizontal line separates the datasets taken during Cycle~2 (until ID 019) from those of Cycle~4. 
}
\label{table:shift}      
\centering                          
\begin{tabular}{c c c r}        
\hline\hline 
ID & $\lambda_0^2$ & RM & $\Delta^{\rm sys} \chi_{\rm ep}$ \\
   & (m$^2$) & (rad m$^{-2}$) & (deg)\\ 
 \hline
009 & 4.4717 & 5.79 $\pm$ 0.01  &  35.0 $\pm$ 3.5 \\
010 & 4.4271 & 5.84 $\pm$ 0.01  &  23.4 $\pm$ 6.1 \\
012 & 4.4226 & 5.87 $\pm$ 0.01  &  4.1 $\pm$ 0.8 \\
013 & 4.3974 & 5.90 $\pm$ 0.01  &  1.5 $\pm$ 1.3 \\
\textbf{014} & 4.4217 & 5.87 $\pm$ 0.01 &  \\
015 & 4.4160 & 5.91 $\pm$ 0.01  &  $-5.0 \pm$ 0.6 \\
016 & 4.4177 & 5.88 $\pm$ 0.01  &  $-3.8 \pm$ 0.6 \\
017 & 4.4177 & 5.83 $\pm$ 0.01  &  17.2 $\pm$ 3.2 \\
018 & 4.4177 & 5.91 $\pm$ 0.01  &  $-11.2 \pm$ 0.5 \\
019 & 4.4228 & 5.84 $\pm$ 0.01  &  20.5 $\pm$ 6.4 \\
\hline
020 & 4.4117 & 5.89 $\pm$ 0.01  &  4.9 $\pm$ 1.2 \\
021 & 4.4026 & 5.90 $\pm$ 0.01  &  $-7.5 \pm$ 0.7 \\
022 & 4.3912 & 5.90 $\pm$ 0.01  &  $-13.0 \pm$ 0.7 \\
023 & 4.3980 & 5.92 $\pm$ 0.01  &  $-15.6 \pm$ 0.6 \\
024 & 4.3713 & 5.92 $\pm$ 0.01  &  $-10.2 \pm$ 1.0 \\
026 & 4.3813 & 5.93 $\pm$ 0.01  &  $-18.5 \pm$ 0.6 \\
028 & 4.4129 & 5.94 $\pm$ 0.01  &  $-17.4 \pm$ 0.5 \\
030 & 4.4129 & 6.06 $\pm$ 0.01  &  $-29.1 \pm$ 0.6 \\
032 & 4.4129 & 6.05 $\pm$ 0.01  &  $-30.1 \pm$ 0.6 \\
\hline                                   
\end{tabular}
\end{table}

For each epoch we computed the difference $\Delta \chi_{\rm{Ep} }^{\rm sys}$ between the polarization angle at the reference epoch (the ``calibrator'') and the polarization angle of the epoch to be corrected. 
The difference $\Delta \chi_{\rm{Ep} }^{\rm sys}$ is calculated as follows:
\begin{equation}
  \Delta \chi_{\rm{Ep} }^{\rm sys} = 
    \chi_{\rm RefEp} - 
   \left( 
  \chi_{\rm{Ep}}+ \rm{RM}_{\rm{Ep}}\cdot(\lambda_{0,\rm{RefEp}}^2-\lambda_{0,\rm{Ep}}^2)
  \right) 
  \, ,
\end{equation}
where  
 $\chi_{\rm {RefEp}}$ is the polarization angle of the reference source at the reference epoch; 
 $\chi_{\rm{Ep}}$ is the polarization angle of the reference source at Epoch Ep; 
 the term $\rm{RM}_{\rm{Ep}}\cdot(\lambda_{0,\rm{RefEp}}^2-\lambda_{0,\rm{Ep}}^2)$ is the rotation due to the Faraday rotation between the different $\lambda_0^2$ values between the reference epoch RefEp and epoch Ep. 
 
The RM values of the reference source ($\rm{RM}_{\rm Ep}$) are listed in the third column of Table~\ref{table:shift}; 
the systematic corrections, $\Delta \chi_{\rm{Ep} }^{\rm sys}$, are given in the last column, for each epoch. 
The errors on the corrections were computed from the standard deviations of $Q(\phi)$ and $U(\phi)$ in the outer 20\% of the Faraday depth range, with sampling 0.3~rad~m$^{-2}$, using error propagation rules.

\subsubsection{Weights}

In the stacking process, weights were attributed to each dataset, based on the noise in each dataset. 
For each epoch, we selected a small ($2.5\times 2.5$~arcmin$^2$) central region, centered on RA\,=\,16$^{\rm h}$11$^{\rm m}$00$^{\rm s}$ and Dec\,=\,55$^{\circ}$00$^{\prime}$00$^{\prime\prime}$, and RM synthesis was performed. The mean of the noise map, $\sigma_{\rm QU,Ep}$, was used to calculate the weight $1/\sigma^2_{\rm QU,Ep}$.
Table ~\ref{table:observations} shows the values of $\sigma_{\rm QU,Ep}$ for each epoch.

\subsubsection{Making strips}

Because of the large size of the datasets (about 400~GB for a single Stokes-parameter frequency cube), we had to divide the field into strips to perform the analysis in parallel on multiple cores of the supercomputer. 
We produced 80 strips of $12005\times 150$ pixels 
(about 5~degree width in RA and 3.75~arcmin height in Dec). The Stokes~$I$ image \citep{Sabater2021} was divided in a similar way. 
All pixels below 320~$\mu$Jy beam$^{-1}$ in total intensity ($\sim$10~$\sigma_{\rm I}$) were masked during RM synthesis.

\subsection{Applying corrections and stacking}

\begin{figure*}[h] 
\includegraphics[width=1.0\linewidth]{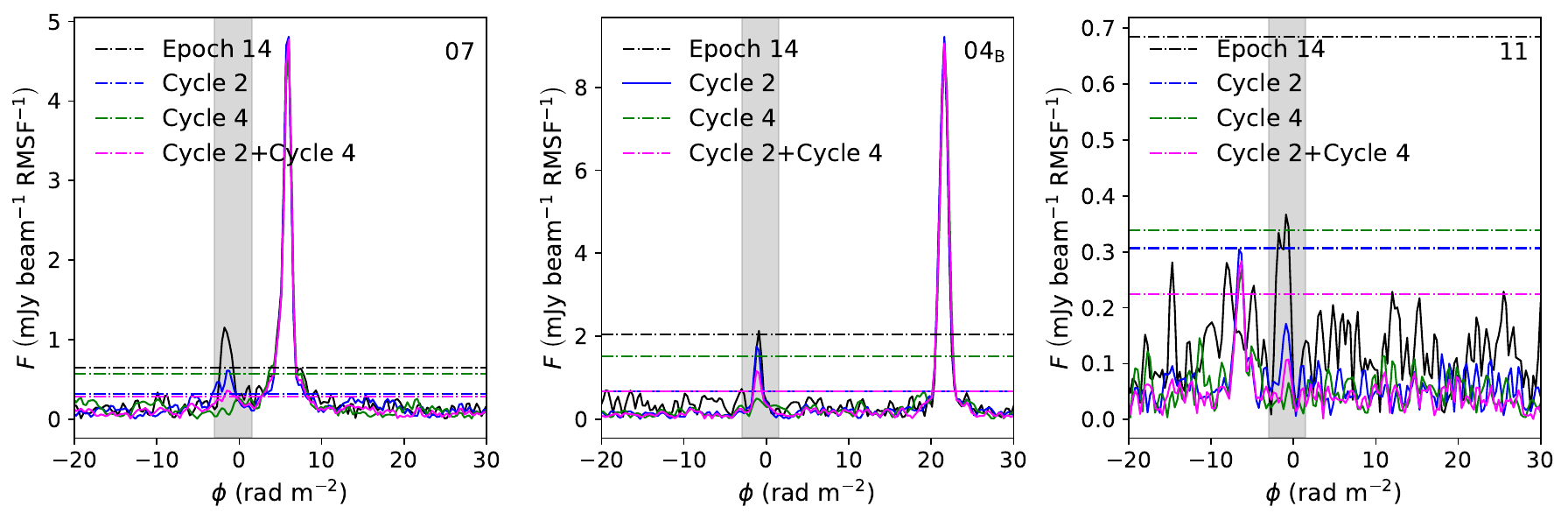}
\caption{
Faraday spectra of three sources: 
the reference source (source 07, left panel); 
the brightest polarized source component (04$_{\rm B}$, middle panel), 
and a faint source detected only after stacking data from both cycles (source 11, right panel). 
The displayed Faraday spectra are from the reference epoch (Epoch~014), from stacked Cycle~2, stacked Cycle~4, and combining both cycles. 
The gray region outlines the range in Faraday depth that was excluded from the analysis due to instrumental polarization. 
The horizontal lines indicate the corresponding $8\sigma_{\rm QU}$ levels. 
Stacking reduces the noise levels (as shown by the magenta horizontal lines that are lowest). The data were of higher quality in Cycle 2 than in Cycle~4 (the horizontal blue lines are below the green ones). 
}
\label{fig:refsource_comp} 
\end{figure*}

We applied the corrections $\Delta \chi_{\rm{Ep}}$ calculated on the reference source 
to the whole field at each epoch:
\begin{equation}
    \mathcal{P}^{\rm{corr}}_{\rm Ep}(i,j,\nu) 
    = \mathcal{P}_{\rm{Ep}}(i,j, \nu)\, 
    \mathrm{e}^{2 {\rm i} \Delta \chi_{\rm{Ep}}}\, .
\end{equation}

Finally, for each cycle, inverse-variance-weighted averages of Stokes $Q$ and $U$ parameters, $Q_{\rm Cycle}$ and $U_{\rm Cycle}$, were calculated at each pixel in the frequency cube: 
\begin{equation}
Q_{\rm Cycle}(i,j,\nu) = 
\dfrac{
        \sum_{\rm Ep} 
           \dfrac{Q_{\rm Ep}^{\rm corr}(i,j,\nu)}
           {\sigma^2_{\rm QU, Ep}}   
           } 
        {\sum_{\rm Ep} 
           \dfrac{1}{\sigma^2_{\rm QU, Ep}} 
           } 
           \, ,
\end{equation}
and similarly for the Stokes parameter $U$. 

Stacked Stokes $Q$ and $U$ cubes of the central region were produced with the same method to compute the weights used to stack the various datasets.

\subsection{Stacking Faraday cubes from different observation cycles} 
\label{stack_2cycle}

Because of the different frequency setup of data taken in Cycle~2 and in Cycle~4, the data could not easily be combined in frequency space but had to be combined in Faraday depth space. 
We performed RM synthesis on each stacked $QU$ frequency cube strip from Cycle~2 and from Cycle~4.

The mean wavelength squared, $\lambda_0^2$, slightly differs between the two cycles: 
\begin{equation} \label{eq:diff}
\Delta \lambda_0^2 =\lambda_{0,\rm{Cycle~2}}^2-\lambda_{0,\rm{Cycle~4}}^2
= 4.475 - 4.287 = 0.188~{\rm m}^2 
\end{equation}
and we used this difference to correct data from Cycle~4:
\begin{equation}
    \mathcal{F}^{\rm corr}_{\rm{Cycle~4}}(i,j,\phi) = 
    \mathcal{F}_{\rm Cycle~4}(i,j,\phi) \,
    \mathrm{e}^{2{\rm i}\phi\Delta \lambda_0^2} \, .
\end{equation}

RM synthesis was performed on the stacked Stokes $Q$ and $U$ cubes of the central region and the means of the noise maps, $\sigma_{\rm QU,Cycle}$, were used to weigh the two cycles. 
The final stacked data cubes (Cycle~2+Cycle~4) of $Q_{\rm st}(\phi)$ and $U_{\rm st}(\phi)$ 
were produced by an inverse-variance weighted average of 
the $Q(\phi)$ and $U(\phi)$ cubes of the two cycles:

\begin{equation}
    Q_{\rm st}(i,j,\phi) = \dfrac{ \dfrac{Q_{\rm Cycle~2}(i,j,\phi)}{\sigma_{\rm QU,Cycle~2}^2} + \dfrac{Q^{\rm corr}_{\rm Cycle~4}(i,j,\phi)}{\sigma_{\rm QU,Cycle~4}^2}}
    {\dfrac{1}{\sigma_{\rm QU,Cycle~2}^2}+\dfrac{1}{\sigma_{\rm QU,Cycle~4}^2}} \, ,
\end{equation}
and similarly for Stokes parameter $U$.

The final Faraday cube F$_{\rm{st}}$($\phi$) was obtained by combining $Q_{\rm st}(\phi)$ and  $U_{\rm st}(\phi)$ at each pixel and $\phi$:  
\begin{equation}
    F_{\rm st}(i,j,\phi)=\sqrt{Q_{\rm st}^2(i,j,\phi)+U_{\rm st}^2(i,j,\phi)}\, ,
\end{equation}
and final maps were produced from the Faraday cubes (polarized intensity map; Stokes $Q$ and $U$ maps; RM map; noise map). 

Figure~\ref{fig:refsource_comp} shows the Faraday spectra of three sources 
in the reference epoch (Epoch~014), in stacked Cycle~2, stacked Cycle~4, and combining both cycles. 
The three sources are the reference source (source 07, left panel), 
source 04$_{\rm{B}}$ (middle panel; 
this is the brightest source in polarized intensity, and the source with the highest RM value), 
and source 11 (right panel; a faint source detectable only after stacking the two cycles). 
These figures illustrate how noise levels in Faraday spectra are decreased by stacking, 
and that the leakage peak is also reduced.
 
\subsection{Noise properties}
\label{sec:noise}

The noise levels, $\sigma_{\rm QU}$, were measured in the central $2.5'\times2.5'$ region, both in datasets from individual epochs 
(Table~\ref{table:observations}) and after stacking (Table~\ref{table:noise_levels_stakeddata}). 
The second column of Table~\ref{table:observations} gives the measured median noise values after stacking, while the third column gives the noise values calculated from Gaussian statistics 
(where $1/\sigma^2 = \sum 1/\sigma_i^2$, where the index $i$ runs from one to the number of elements in the stacked dataset). 
The decrease in the noise level after stacking is in agreement with expectations from Gaussian statistics. 
\begin{table}[h]
\caption{Measured noise levels in the central $2.5'\times2.5'$ region after stacking and expected values from Gaussian statistics.} 
\centering                         
\label{table:noise_levels_stakeddata}   
\begin{tabular}{ c  c  c}        
\hline\hline                
Dataset & Measured $\sigma_{\rm{QU}}$ & Expected $\sigma_{\rm{QU}}$ \\ 
        & ($\mu$Jy~beam$^{-1}$) & ($\mu$Jy~beam$^{-1}$) \\
\hline
Cycle~2 & 30  & 29 \\
Cycle~4 & 32  & 30 \\
Cycle~2+Cycle~4 & 22  & 20 \\
\hline                                   
\end{tabular}
\end{table}

\begin{figure}[h] 
\includegraphics[width=\linewidth]{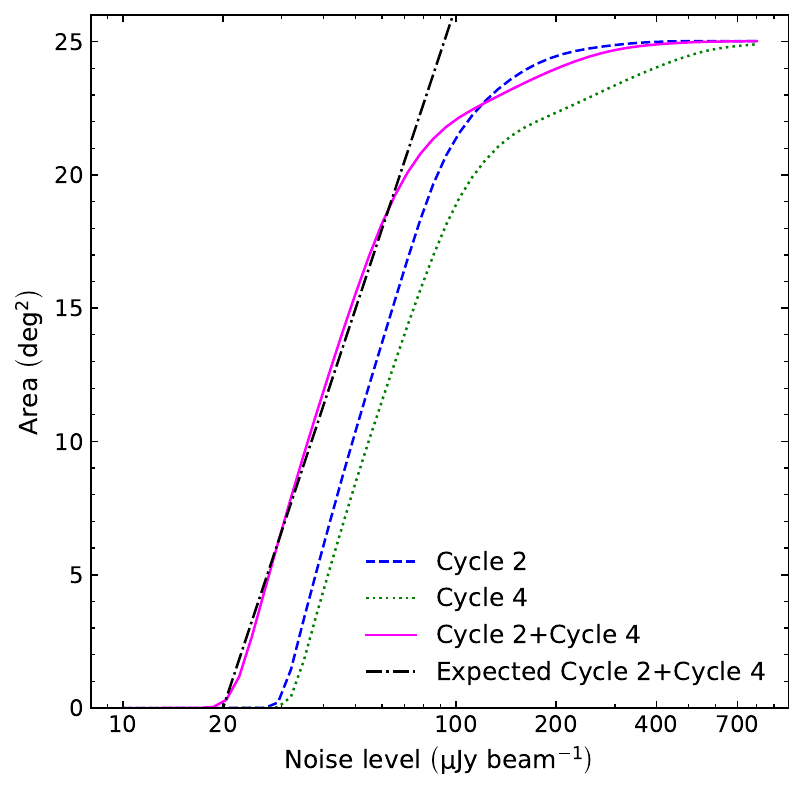} 
\caption{Area of the field with noise $\sigma_{\rm QU}$ equal to or lower than a given value in data from stacked individual cycles and in the final dataset. The black dashed curve is the expected profile from a Gaussian primary beam (eq.~\ref{eqAreavsNoise}), with $\sigma_0 = 20~\mu$Jy~beam$^{-1}$.
}
\label{fig:noise_area}
\end{figure}

Figure~\ref{fig:noise_area} shows the improvement in the sensitivity brought by stacking: the noise level in the field is lower after stacking the datasets from both cycles. 
The curves also clearly show the better quality of the  Cycle~2 data compared to Cycle~4 data. 
Up to a radius of 2.5$^\circ$ (which corresponds to an area of 19.6 deg$^2$), the shape of the curves follows the profile expected from the noise increase with radius within a circular Gaussian primary beam, in which case the corresponding area with a noise lower than a given $\sigma$ is given by: 
\begin{equation}
A ({\rm noise} < \sigma) = 
\frac{\pi}{4 \ln 2} \theta_{\rm FWHM}^2 \ln\left(
\frac{\sigma}{\sigma_0} \right) ,
\label{eqAreavsNoise}
\end{equation}
where $\theta_{\rm FWHM} = 3.80^\circ$ 
is the full-width half maximum of the primary beam of the LOFAR High Band Array at 150~MHz\footnote{\url{https://science.astron.nl/telescopes/lofar/lofar-system-overview/observing-modes/lofar-imaging-capabilities-and-sensitivity/}}
and $\sigma_0$ is the value of the noise in the center of the image, as shown by the dashed black curve. 
The noise increases rapidly for areas greater than about 20 deg$^2$; this is because of the contributions from pixels in the four corners of the map in the regions 
that are within the $5\times 5$ deg$^2$ area but outside the disk of 2.5$^\circ$ radius. 

\vfill\eject
\section{Source finding} \label{sec:source_finding}

The analysis was performed for the whole 25~deg$^2$ field, where a threshold of 8$\sigma_{\rm QU}$ was used. A lower threshold of 6$\sigma_{\rm QU}$ was used for the sources known to be polarized at 1.4~GHz (either from the DRAO work of \citealt{grant2010} or from the NVSS RM catalog of \citealt{taylor2009}). 

\subsection{Whole field} 
\label{sec:analysis_allfield}

As described in Sect.~\ref{sec:stacking}, the large data cubes were divided into strips, and datasets from all epochs were stacked. 
We created S/N maps by dividing the polarized intensity map of each strip by its corresponding noise map. 
Pixels with S/N~$> 8$ were selected, as they represent the most reliable detections: indeed, \cite{George2012} found the false detection rate to be less than 10$^{-4}$ for an 8$\sigma_{\rm{QU}}$ detection threshold.

A preliminary catalog was then made by cross-matching the positions of the detections with the catalog of Stokes $I$ components in the field from \cite{Sabater2021}. We used a cross-matching radius of 1~arcmin in order to include matches with extended sources.
This resulted in a preliminary catalog of 84 matches. Because of the size of the cross-matching radius, some matched Stokes~$I$ components (including artifacts) referred to the same detection. We selected the Stokes~$I$ components that were counterparts of the detections, resulting in a list of 44 polarized components. We identified two regions of diffuse Stokes~$I$ emission that had been cataloged as multiple Stokes~$I$ components by \cite{Sabater2021}; for these, we selected the component closer to the polarization detection. 

\subsection{Sources known to be polarized at 1.4 GHz}  \label{sec:ident_1.4}

This analysis was performed on sources known to be polarized at 1.4~GHz. 
We selected the pixels with S/N $> 6$ in each strip and cross-matched this list with the DRAO ELAIS-N1 catalog of \cite{grant2010} and the NVSS~RM~catalog of \cite{taylor2009}, using a conservative radius of 1.5~arcmin.
This resulted in the identification of 21 components between 6 and $8\sigma_{\rm QU}$ from the DRAO ELAIS-N1 catalog, but none from the NVSS RM catalog. 

\subsection{Identification of the most reliable sources} 
\label{sec:reliable}

For each potentially polarized source, we visually inspected the LOFAR Stokes~$I$ image, the polarized intensity image (POLI), and the Faraday spectrum. The following criteria were used to reject candidates from the final catalog: 
\begin{enumerate}
\item only one isolated pixel in the POLI image was found above the threshold, or no Stokes~$I$ counterpart was associated with a group of pixels above the threshold in POLI; 

\item the POLI detection was associated with image artifacts (for instance around bright Stokes $I$ sources);

\item the Faraday spectrum showed multiple peaks close to the instrumental polarization region and/or in the outer part of the spectrum;

\item the fractional polarization was greater than 100\% or lower than 0.2\% (this lower limit on fractional polarization was also used by \citealt{HerreraRuiz2021}).
\end{enumerate}

This resulted in the rejection of: 

-- 19 sources from the whole field above $8\sigma_{\rm QU}$
(out of which 15 were in the DRAO ELAIS-N1 catalog of polarized sources and 7 in the NVSS RM catalog);  

-- 29 sources detected at 1.4~GHz by \cite{grant2010} 
(14 in the 6--8$\sigma_{\rm QU}$ range 
and 15 above 8$\sigma_{\rm QU}$). 

\section{Result and discussion} 
\label{sec:results}

In this section we start by presenting the list of polarized sources identified in the field. 
We compare our results with previous works on polarization. 

\subsection{Catalog of polarized sources} \label{sec:catalog}

In Table~\ref{table:det_cat} we provide a list of the detected polarized components. 
The results can be summarized as follows: 
\begin{itemize} 
    \item [--] The first 26 entries (above the horizontal line) correspond to polarized components detected above 8$\sigma_{\rm QU}$ in the 25~deg$^2$ field; of these, six components were detected in the reference epoch alone; 16 were also detected in the stacked Cycle~2 data, and 15 in the stacked Cycle~4 data; 13 were detected in polarization at 1.4~GHz by \cite{grant2010}. 
    \item [--] The next seven entries below the horizontal line correspond to polarized components that were known to be polarized at 1.4~GHz from the \cite{grant2010} study and in which we searched for polarization in the 6--8$\sigma_{\rm QU}$ range of the LOFAR data. Of these, two were detected in stacked Cycle~2 data and three in stacked Cycle~4 data. One of the entries (29) corresponds to the second lobe of Source~13. 
    \item[--] The 33 polarized components detected by stacking Cycle~2 and Cycle~4 data are therefore associated to 31 radio continuum sources: in one radio galaxy, polarization is detected from both lobes above 8$\sigma_{\rm QU}$ (components $04_{\rm A}$ and $04_{\rm B}$), 
    while in another radio galaxy, one lobe is polarized above 8$\sigma_{\rm QU}$ (component~13) 
    and the other lobe below 8$\sigma_{\rm QU}$ (component~29). 
    \item[--] The polarized intensities (in column 5) were corrected for the polarization bias following eq.~5 from \cite{George2012}:
 \begin{equation}
     P = \sqrt{\Tilde{P}^2 - 2.3 \sigma^2_{\rm QU}} \, ,
 \end{equation}
 where $\Tilde{P}$ is the polarized intensity before correction, from the $F_{\rm {map}}$, 
 and $\sigma_{\rm QU}$ is the mean of the rms noise in $Q(\phi)$ and $U(\phi)$, calculated in the outer 20\% of the Faraday depth range. 
\item [--] The RM values of the sources and their corresponding uncertainties are given in column~6. 
A parabola was fitted to the polarized intensity peak in the Faraday spectra to improve the precision on the RM measurements. The error in the RM was calculated as the RM resolution, $\delta \phi$,  divided by twice the S/N of the detection, following \cite{Brentjens2005}. 
 \end{itemize} 

\subsection{Fraction of missed polarized sources} \label{sec:missing_sources}
A number of polarized sources is missing from the catalog for at least two reasons: the exclusion of part of the RM range due to instrumental polarization, and the threshold that we imposed on the minimal fractional polarization, as discussed below.

Because of instrumental polarization due to leakage in the RM range $[-3,1.5]$~rad~m$^{-2}$, this range in the Faraday cubes was excluded from the search (Sect.~\ref{sec:RMsyth}). Assuming a uniform distribution between the lowest and highest RM values of our catalog, we estimate that about four polarized sources were missed.

The threshold we imposed on the minimal reliable fractional polarization, equal to 0.2\% based on the study of \cite{HerreraRuiz2021}, may have caused us to miss the detection of real polarized sources. 
In the LoTSS-DR2 RM grid catalog, such a threshold was not applied and \cite{lotssdr2} found that $\sim$3\% of polarized sources have a degree of polarization lower than 0.2\% (and above 0.05\%). 
We detected 25 sources above 8$\sigma_{\rm{QU}}$ after stacking, and may be missing $\sim$1 polarized source because of the threshold in the minimal fractional polarization. We re-analyzed the preliminary list of 44 components detected above 8$\sigma_{\rm QU}$ removing the threshold in the minimal fractional polarization and we found one possible candidate as polarized source. We decided, however, not to include this source because the region of the Faraday spectrum close the detection, at $\sim$5~rad~m$^{-2}$, showed several peaks close to the 8$\sigma_{\rm QU}$ threshold, making the detection uncertain.

We therefore estimate that about five sources were missed; taking this into account, the number of polarized sources above 8$\sigma_{\rm QU}$ would be $\sim30$ instead of 25 in the 25~deg$^2$ region of the LOFAR ELAIS-N1 field. 

\subsection{Comparison with previous polarization studies of the ELAIS-N1 LOFAR deep field}
\label{sect:comp_noelia}

\begin{table}[h]
\caption{RM values from LOFAR at 
6"-resolution (this work), 20"-resolution \citep{HerreraRuiz2021} and NVSS \citep{taylor2009}.} 
\label{table:rm_comp}      
\begin{tabular}{ c r r r }        
\hline\hline                
Source ID & RM$_{6"}$  & RM$_{20"}$ &  RM$_{\rm NVSS}$ \\ 
& (rad m$^{-2}$) & (rad m$^{-2}$) & (rad m$^{-2}$)\\
 \hline
03   & -5.83 $\pm$ 0.02 &  -5.80 $\pm$ 0.03 & - \\
04$_{\rm B}$ & 21.78 $\pm$ 0.01 & -  & 20.7 $\pm$ 7.0  \\
06   & 18.43 $\pm$ 0.03 &  18.44 $\pm$ 0.04 &   -  \\
07   & 6.062 $\pm$ 0.003 &  6.12 $\pm$ 0.01 &   -0.9 $\pm$ 7.5  \\
12   & 7.17 $\pm$ 0.02 &  7.30 $\pm$ 0.02 &    -  \\
14   & 10.31 $\pm$ 0.02 &  10.39 $\pm$ 0.04 &    - \\
15   & -4.80 $\pm$ 0.04 &  -4.86 $\pm$ 0.05 &   -  \\
17   & 1.95 $\pm$ 0.03 &  1.43 $\pm$ 0.04 &    - \\
18   & -20.31 $\pm$ 0.03 &  -20.15 $\pm$ 0.03 &    -   \\ 
10   & 2.91 $\pm$ 0.04  & - & -17.3 $\pm$ 12.4  \\
20   & -4.70 $\pm$ 0.02 &  -4.70 $\pm$ 0.05 &    - \\
24   & 9.43 $\pm$ 0.01 &  9.46 $\pm$ 0.01 &  -7.1 $\pm$ 4.2  \\
25   & 10.03 $\pm$ 0.02  &  -   &  -2.8 $\pm$ 8.4 \\
\hline                                   
\end{tabular}
\end{table}

\begin{figure}[h] 
\centering
\includegraphics[width=8.8cm]{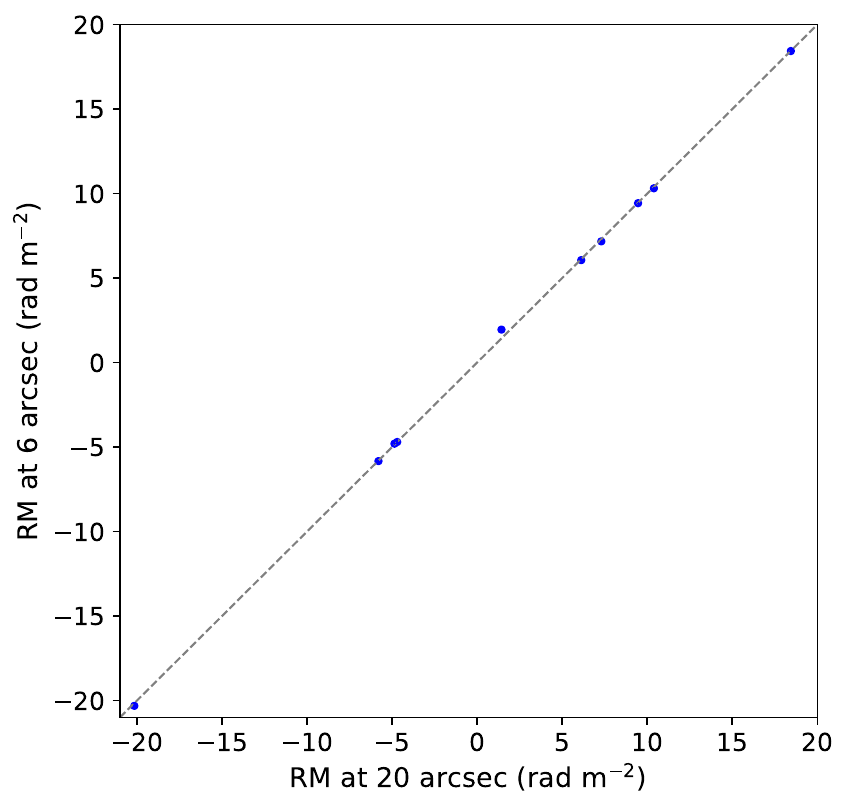}
\includegraphics[width=8.8cm]{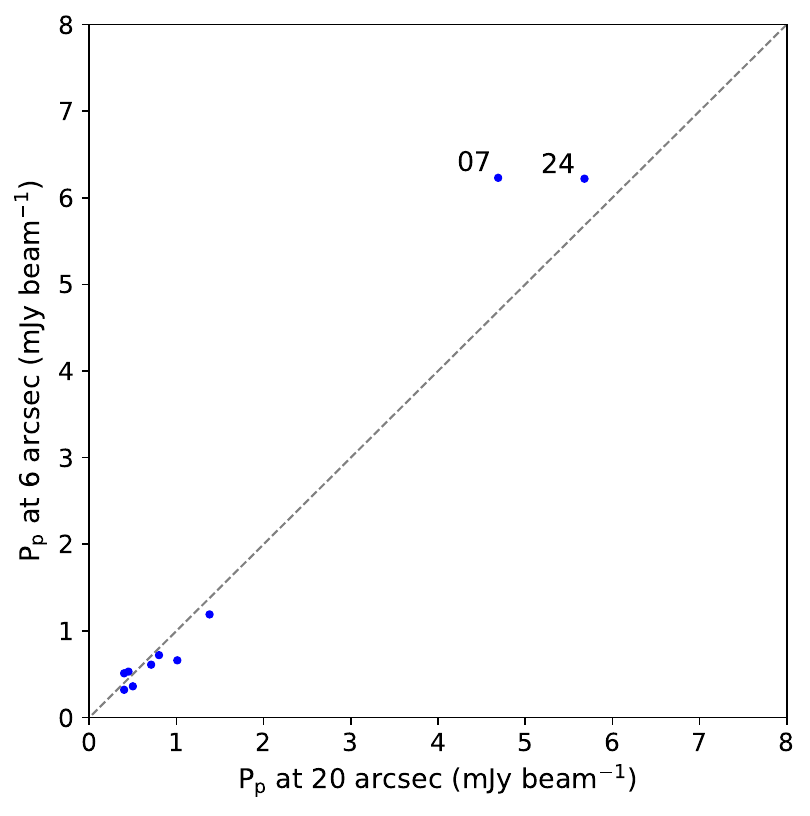} 
\caption{
Comparison of RM values and peak polarized intensities measured with LOFAR at 6" resolution (this work) 
and at 20" resolution for the ten sources of \cite{HerreraRuiz2021}. 
The dashed lines are the lines of equality.
}
\label{fig:comp_6_20as}
\end{figure}

\subsubsection{20" resolution}
In their pilot study of polarization in the ELAIS-N1 LOFAR deep field, \cite{HerreraRuiz2021} 
detected ten polarized sources in a 16-deg$^2$ field imaged at a resolution of 20$''$.
All those sources are detected in the deeper, $6''$-resolution data presented here. 
The RM values agree, as shown in the left panel of Fig.~\ref{fig:comp_6_20as} and in Table~\ref{table:rm_comp}. 
The peak polarized intensities at $20\arcsec$ and $6\arcsec$ are in agreement, too (right panel of Fig.~\ref{fig:comp_6_20as}), except for the brightest polarized sources, Source~07 (the reference source) and Source~24, where the peak polarized intensities are higher at 6$''$ than at 20$''$. 
The reference source is a double-lobed radio galaxy (Sect.~\ref{sec:refsource}), with two components separated by $\sim12''$, and the different levels of polarization are probably due to beam depolarization within the 20$''$ beam. For Source~24 (a blazar), the polarized intensity at $6''$ is higher than at 20$''$ by about 10 percent; this can be due to calibration uncertainty or to the variability of the source.

\subsubsection{4.3' resolution}
\cite{Snidaric2023} carried out a deep polarimetric study of Galactic synchrotron emission at low radio frequencies by stacking 21 epochs of the ELAIS-N1 LOFAR Deep Field at 4.3 arcmin-resolution. They also checked how many extragalactic polarized sources from the catalog of \cite{HerreraRuiz2021} were detected at their resolution, and found nine out of ten radio sources (our IDs: 03, 06, 07, 12, 14, 17, 18, 20, 24).
By inspecting the Faraday spectra of the stacked polarized intensity cube of \cite{Snidaric2023} at the locations of our polarized sources, we could see that five additional sources (our IDs: 04, 10, 11, 15, 25) were visible in the low-resolution polarization data, while the others were contaminated by diffuse polarized emission. 
The rotation measures of the detected sources are in agreement despite the very different angular resolutions of the data. 

\subsection{Comparison with NVSS 1.4~GHz RM catalog}
The NVSS 1.4~GHz RM catalog \citep{taylor2009} contains entries for 25 radio sources in the ELAIS-N1 25~deg$^2$ field. 
Of these 25 sources, five are in our LOFAR catalog and three 
of those are the top brightest in polarization (sources~04, 07, 24). 
Table~\ref{table:rm_comp} lists the RM values from LOFAR and from NVSS for the five 
matches. 
They are in agreement within the relatively large uncertainties on the RM values from NVSS (several rad~m$^{-2}$). 
The different RM values for Source~24 may come from the fact that the source is a blazar that may be variable (\citealt{Anderson2019}; \citealt{HerreraRuiz2021}). 

\subsection{Depolarization} \label{sec:comp_1.4}

\begin{table}[h]
\centering
\caption{Stokes $I$ intensities (in mJy~beam$^{-1}$) and corresponding degrees of polarization (in $\%$) at the pixel of peak polarized intensity for the 20 polarized sources detected both at 150~MHz (this work) and at 1.4~GHz \citep{grant2010}.}  
\label{table:1.4_comp}      
\begin{tabular}{ c r r r r r}        
\hline\hline                
Source ID & $I_{\rm p,150}$  &  $I_{\rm p,1420}$ & $\Pi_{150}$ & $\Pi_{1420}$ & $\Delta$pos\\ 
 \hline
$05^{a}$   & 39.30 & 43.70 &  0.86 & 5.24  &22.1\\
07   & 355.62 & 218.97 &  1.75 & 4.68 &5.7\\
08   & 48.38 & 13.53 &  0.82 & 11.62 &11.9\\
10   & 54.61 & 262.79 &  0.45 & 1.41 &1.3\\
11   & 38.32 & 28.01 &  0.75 & 6.66  &10.8\\
12   & 18.68 & 43.84 &  3.86 & 1.47  &7.0\\
$15^{b}$   & 2.18 & 5.23 &  14.76 & 14.86  &81.0\\
16   & 3.85 & 6.98 &  5.17 & 10.01   &8.0\\
18   & 9.00 & 12.10 &  4.07 & 8.03   &2.4\\
$19^{c}$   & 39.22 & 58.09 &  0.77 & 8.04  &18.2\\
20   & 33.55 & 55.89 &  1.52 & 3.76  &3.4\\
23   & 15.10 & 8.04 &  2.39 & 9.57   &4.6\\
25   & 71.80 & 85.47 &  1.35 & 7.94  &0.9\\
26   & 10.17 & 11.99 &  1.55 & 8.43  &13.1\\
$27^{d}$   & 3.18 & 44.81 &  7.02 & 2.08   &33.1\\
28   & 36.72 & 22.87 &  0.56 & 8.35  &1.6\\
29   & 8.62 & 5.17 &  2.19 & 12.47   &8.1\\
$30^{e}$   & 29.41 & 53.21 &  1.11 & 14.64 &14.3\\
31   & 18.50 & 60.76 &  1.32 & 4.14  &0.8\\
32   & 23.05 & 38.61 &  0.82 & 3.00  &12.8\\
\hline                                   
\end{tabular}
\tablefoot{The last column gives the offset in arcseconds between the peak polarized intensity in the 1.4~GHz map \citep{grant2010} and in the LOFAR map (this work).\\
$^{a}$ In Source 05, the peak polarization at 150~MHz and at 1.4~GHz is detected from different radio lobes on either side of the nucleus; \\
$^{b}$ In Source~15 the polarization peaks on different radio components at 1.4~GHz and at 150~MHz; \\
$^{c}$ In Source~19 the peak polarization at 1.4~GHz is found close to the center while it is in a lobe at 150~MHz;\\
$^{d}$ In Source~27 the peak polarization at 1.4~GHz is found close to the center while it is in a lobe at 150~MHz; \\
$^{e}$ In Source~30 the peak polarization is close to the center while it is in a lobe at 150~MHz. 
}
\end{table}

\begin{figure*}[h] 
\centering
\includegraphics[width=0.49\linewidth]{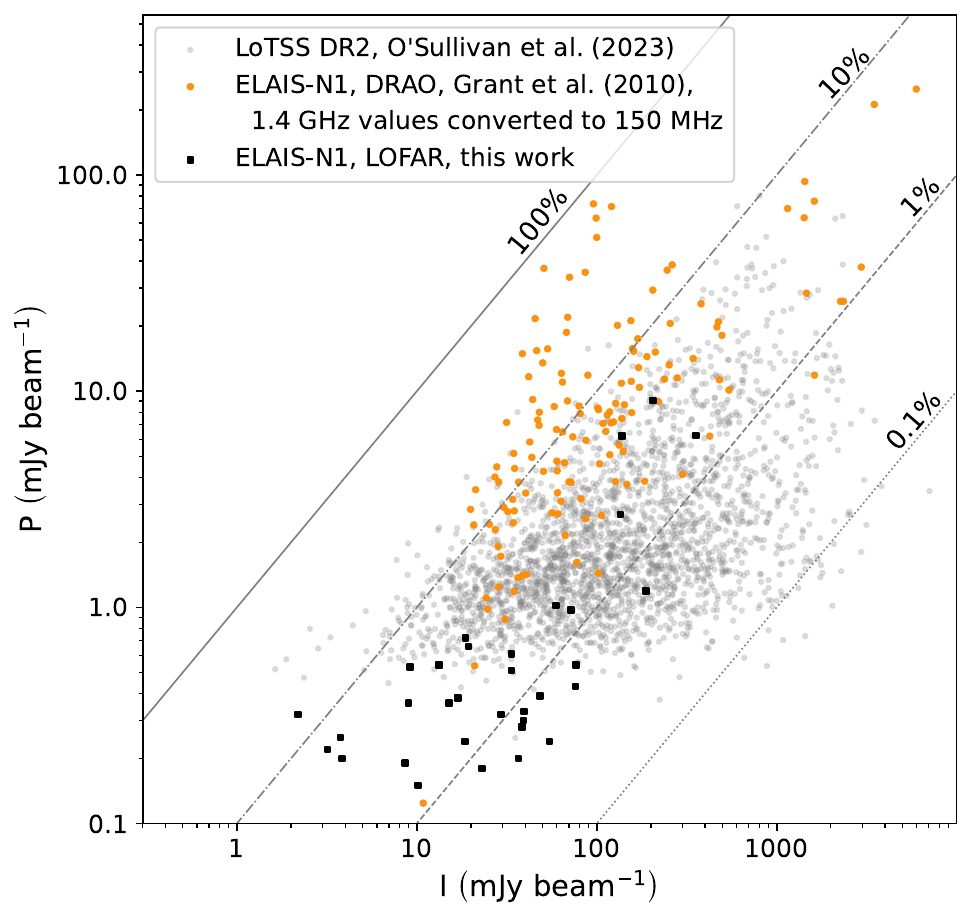}
\includegraphics[width=0.49\linewidth]{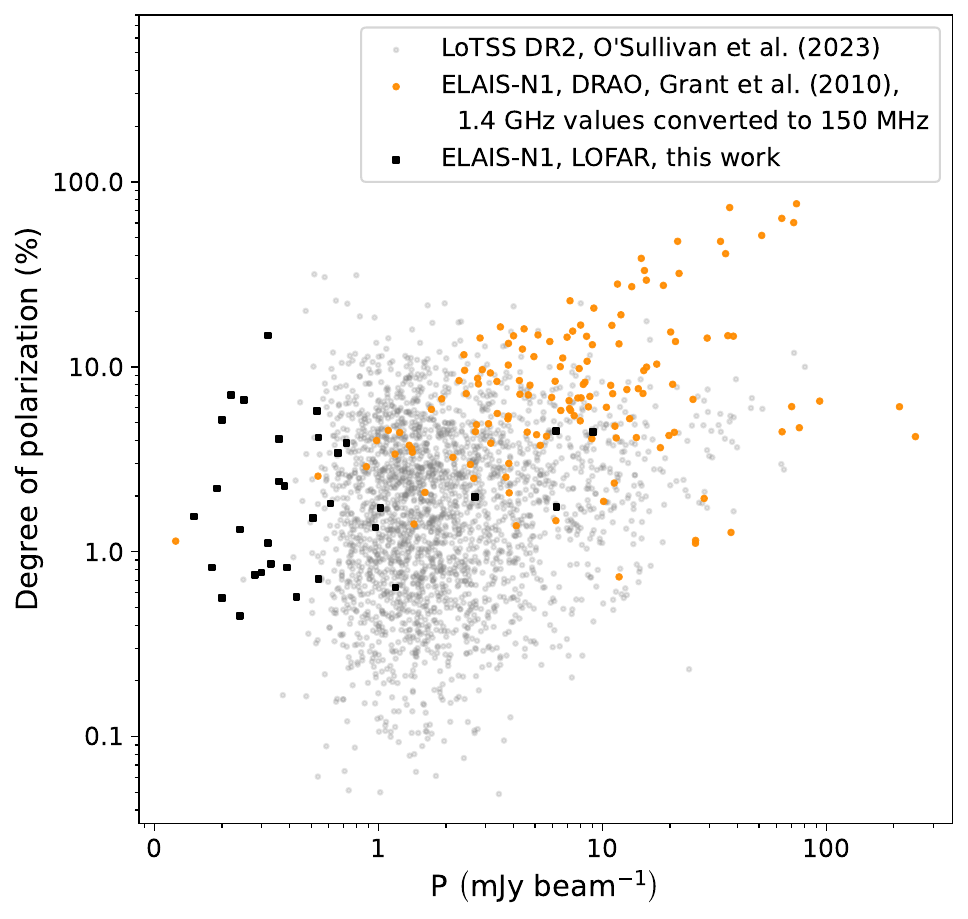} \\
\caption{
Peak polarized intensity vs total intensity (left panel) 
and degree of polarization vs peak polarized intensity (right panel)
for three catalogs:  
in gray the LoTSS-DR2 RM grid catalog of \cite{lotssdr2}; 
in black the polarized sources in ELAIS-N1 at 150~MHz (this work); 
in orange the values for sources in the DRAO ELAIS-N1 catalog, 
where both $I$ and $P$ have been converted from 1.4~GHz to 150~MHz using the spectral indices given by \cite{grant2010}. 
The diagonal lines correspond to the indicated degrees of polarization. 
}
\label{fig:polVSsi}
\end{figure*}

In Fig.~\ref{fig:polVSsi} we show the distributions of peak polarized intensities, corresponding total intensities and fractional polarizations for the polarized sources in the LoTSS-DR2 RM grid catalog (20$''$ resolution, \citealt{lotssdr2}), 
our deep LOFAR catalog of polarized sources in ELAIS-N1 (6$''$ resolution), 
and the polarized sources in the DRAO ELAIS-N1 catalog at 1.4 GHz \citep{grant2010}. 
Our study reaches lower levels in polarized intensities ($\leq 0.7$~mJy~beam$^{-1}$) than the shallower but much larger LoTSS-DR2 RM grid.
In terms of degrees of polarization, however, they do not appear to be very different from those of sources in the LoTSS-DR2 RM catalog.

Our LOFAR catalog of polarized sources and the DRAO ELAIS-N1 Source Catalog make it possible to investigate the depolarization between 150~MHz and 1.4~GHz for the sources in common. 
In Table~\ref{table:1.4_comp} the names of these 20 sources are listed, along with the values of their total intensity and degree of polarization at both frequencies. In the last column we give the positional offsets of the peak polarized intensities at 1.4~GHz \citep{grant2010} and 150~MHz (this work). 
In three sources (15, 27, 05), the offsets are greater than 20$''$: in Source~15 (offset of $81''$) and Source~05 the polarization was found in different lobes, while in Source~27 the peak polarization was from the center at 1.4~GHz but from a lobe at 150~MHz. Also in Sources~19 and 30, the polarization peaks near the center at 1.4~GHz, but in a lobe at 150~MHz. These differences are likely due to stronger depolarization at LOFAR frequencies that favors detection of polarization from outer regions of radio galaxies that are less depolarized by the magneto-ionic medium of the host galaxies.  

\begin{figure}
\includegraphics[width=\linewidth]{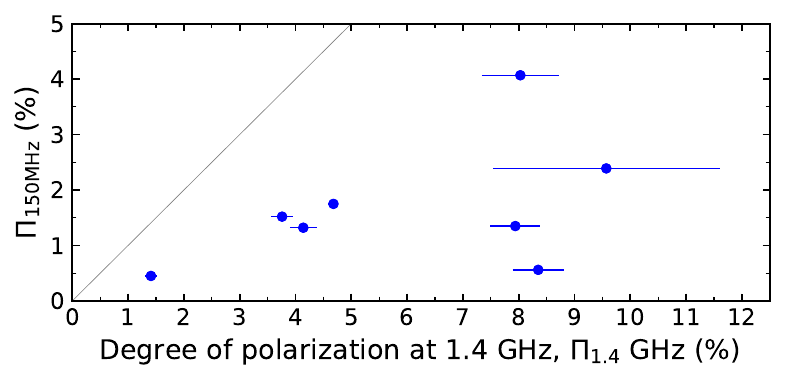} 
\caption{Degrees of polarization at 150~MHz and 6$''$ resolution (this work) and 1.4~GHz and $49''\times49''/\sin\delta$ resolution \citep{grant2010} 
for the eight sources of ELAIS-N1 with both LOFAR and DRAO polarization data 
and with peak polarized intensities that coincide within $6''$. 
}
\label{fig:comp_150_1.4}
\end{figure}

In Fig.~\ref{fig:comp_150_1.4} we show the fractional polarization at 150~MHz versus that at 1.4~GHz for the eight sources whose peak polarized intensity at 1.4~GHz and at 150~MHz coincide within $6''$ (the resolution of the LOFAR observations; the uncertainties given by \citealt{grant2010} on the positions of the peak polarized intensities at 1.4~GHz are of the order of $1''$). 
All sources show a lower fractional polarization at lower frequencies. The fractional polarization is calculated by dividing the peak polarized intensity by the value of the Stokes $I$ intensity at the corresponding pixel. Despite the larger beam at 1.4~GHz that covers more of the extended Stokes $I$ emission, all sources show significant depolarization at 150~MHz, an indication of Faraday depolarization across the 6$''$ beam. 

\subsection{Number counts}

Source counts in polarization are unknown at low radio frequencies as most analyses were carried out at 1.4~GHz 
\citep{Tucci2004, BeckGaensler2004, taylor2007, OSullivan2008, Hales2014_440, Berger2021}. 
\cite{Tucci2004} investigated the distribution of fractional polarization for NVSS sources with $S > 100$~mJy. 
\cite{BeckGaensler2004} modeled the distribution of fractional polarization for NVSS sources with total flux density $S > 80$~mJy, using two log-normal distributions.
\cite{taylor2007} studied the distribution of fractional polarization of fainter sources  ($S < 30$~mJy) in the ELAIS-N1 field and found a constant fractional polarization for sources brighter than 0.5~mJy, where the fainter source population is more strongly polarized. 
\cite{Stil2014} also observed an increase of median fractional polarization for the fainter sources in the stacked NVSS polarization data, but more gradual than what was found in previous polarization studies. 
In the studies of their fields, \cite{Hales2014_440} and \cite{Berger2021} found this to be a selection effect.
\cite{OSullivan2008} simulated the extragalactic polarized sky by using Gaussian fractional polarization distributions and a semi-empirical simulation of the extragalactic total intensity continuum sky \citep{Wilman2008}, finding a good agreement with the observational results from the NVSS. 

\begin{table*}[h]
\caption{Polarized source counts in ELAIS-N1 at 150~MHz.}  
\centering                         
\label{table:counts}      
\begin{tabular}{ c  c  c  c  c c c}        
\hline\hline  
Bin & $P$       & $N$   & $N_{\rm corr}$ & $\Omega_{\rm eff}$ & $n(P) P^{2.5}$ &$n_{\rm corr}(P) P^{2.5}$\\
(mJy) & (mJy)   &       &               & (deg$^2$) & (Jy$^{1.5}$ sr$^{-1}$) & (Jy$^{1.5}$ sr$^{-1}$)\\
(1) &(2)        &(3)    &(4)    &(5)    &(6) &(7)\\
\hline
0.2--0.5  & 0.316  &12  & 14.4 & 8.677  & $(2.69 \pm 0.94)\cdot 10^{-2}$ &$(3.23 \pm 1.13)\cdot 10^{-2}$ \\ 
0.5--1.0  & 0.707  & 8   & 9.6 & 20.211 & $(3.46 \pm 1.22)\cdot 10^{-2}$ &$(4.15 \pm 1.47)\cdot 10^{-2}$ \\ 
1.0--12.0 & 3.464  & 5   & 6   & 24.599 & $(4.28 \pm 1.92)\cdot 10^{-2}$ &$(5.14 \pm 2.30)\cdot 10^{-2}$ \\ 
\hline                                   
\end{tabular}
\tablefoot{Range of polarized flux densities (column 1) and value of $P$ corresponding to the middle of the bin in logarithmic space (column 2). Counts based on the catalog of 25 polarized sources detected above 8$\sigma_{\rm QU}$ 
(columns 3 and 6) 
and corrected to take into account the missing sources (columns 4 and 7; total of 30 sources). 
The fifth column gives the effective areas used to calculated the number densities of sources in each bin. 
In the case of Source~4, the polarized flux densities from both components were added to place it in a flux-density bin, but the mean of the polarized flux densities was used to select the area within which the source could be detected.}
\end{table*}

We estimated the number counts from the catalog of polarized sources detected above 8$\sigma_{\rm QU}$ as this catalog is well defined and complete in the 25~deg$^2$ area of the LOFAR ELAIS-N1 field, while the catalog of sources detected in the 6-to-8$\sigma_{\rm QU}$ range contains only sources that had a prior detection in the 1.4~GHz catalog of \cite{grant2010}.  
All our detections are point-like in polarization, so the peak polarized intensity values (in, for instance, $\mu$Jy~beam$^{-1}$) correspond to the polarized flux densities (in $\mu$Jy). 

In Table~\ref{table:counts} we present the derived polarized source counts for the LOFAR ELAIS-N1 field. 
The data were split into three bins of polarized flux densities. The actual numbers of sources detected in each bin are given in the third column. 
To calculate the effective area of the field, $\Omega_{\rm eff}$, within which the  polarized sources in a given bin could be detected at S/N~$\geq 8$ (Fig.~\ref{fig:noise_area}), we followed \cite{condon1982}: 
the weighted number of sources per square degree, $N_w$, in each flux-density bin can be obtained from 
\begin{equation}
N_w = \sum_{i=1}^{N} \frac{1}{\Omega_i} = \frac{N}{\Omega_{\rm eff}}\, , 
\end{equation}
where $\Omega_i$ is the area within which the $i$th source in the considered bin could be detected and $N$ is the total number of sources in that bin. The errors on the number counts were propagated from the statistical error in $N_w$, given by  
\begin{equation}
 \sigma_{N_w} = \left( \sum_{i=1}^{N} \frac{1}{\Omega_i} \right)^{1/2} \, .
\end{equation}

\begin{figure}[h]
\includegraphics[width=\linewidth]{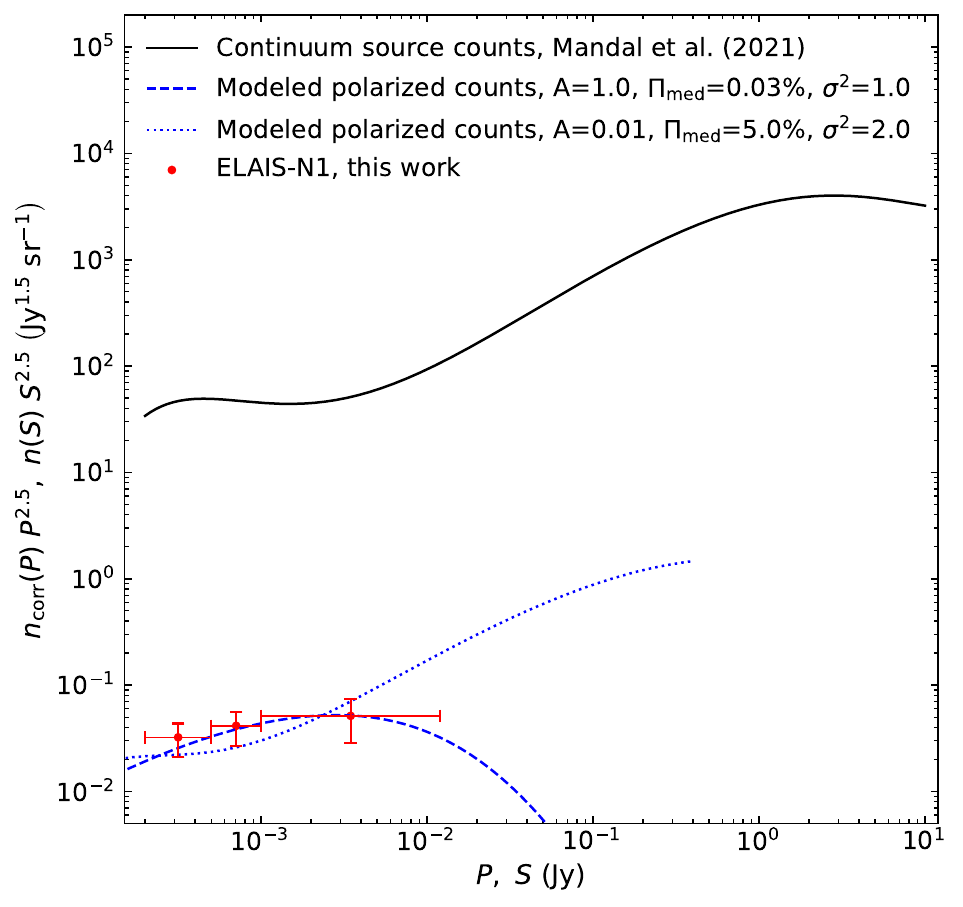}
\caption{Euclidean-normalized differential source counts at 150~MHz in radio continuum and in polarization.
The solid line shows the fit to the source counts in Stokes $I$ in three LOFAR deeps fields \citep{Mandal2021}. 
The other lines show the modeled polarized source counts in ELAIS-N1. 
The red dots represent our data;  
the horizontal bars indicate the widths of the bins. }
\label{fig:counts} 
\end{figure}

The counts were corrected to take into account the five missed sources (discussed in Sect.~\ref{sec:missing_sources}), 
by scaling the number in each bin by a factor of 1.2, equal to the ratio of the expected total number of sources over $8 \sigma_{\rm QU}$ to the number of detected sources, 30/25. 
The last columns give the Euclidean-normalized differential source counts before and after correction for the missed sources; the values for the corrected counts are shown as red dots in Fig.~\ref{fig:counts}. 
Also shown on Fig.~\ref{fig:counts} are the modeled polarized source counts, which were computed as described below. 

The differential polarized source counts $n(P)$ can be obtained by convolving the sources counts in Stokes $I$, $n(S)$, with the probability distribution function of the fractional polarization, $\mathcal{P}(\Pi)$
(e.g. \citealt{BeckGaensler2004}, \citealt{Tucci2012}), 
assuming $\Pi$ independent of total flux density $S$ and for sources with $S \geq S_0$:
\begin{equation}
    n(P)= A\int_{S_0=P}^{\infty} \mathcal{P} \left( \Pi=\frac{P}{S} \right) \,  n(S) \, \frac{dS}{S} \, ,
    \label{eq_nofP}
\end{equation}
where $A$ is a scaling factor. 
As for $n(S)$, we use the empirical polynomial fit performed by \cite{Mandal2021} (their equation 13) to the deepest source counts in Stokes $I$ to date at 150~MHz: the ones based on the 6$''$-resolution images from the central parts of the LoTSS Deep Fields (10.3 deg$^2$ in the Lockman Hole, 8.6~deg$^2$ in Boötes, and 6.7~deg$^2$ in ELAIS-N1; \citealt{Tasse2021}, \citealt{Sabater2021}). 

The probability distribution function of fractional polarization is taken to be a log-normal function:
\begin{equation}
    \mathcal{P}(\Pi) = \frac{1}{\sigma \sqrt{2 \pi}\Pi} 
    \exp \left(   
     - \frac{  \Bigl[ 
    \log (  \Pi / \Pi_{\rm med} ) 
    \Bigr]^2}
    {2 \sigma^2}
    \right) \, ,
\end{equation} 
where $\Pi_{\rm med}$ is the median of the distribution and 
$\sigma^2 = \frac{1}{2}  \log{ \left( \langle \Pi^2 \rangle / \Pi_{\rm med}^2 \right)}$. 

The model of polarized source counts has three parameters: $\Pi_{\rm med}$, $\sigma$, and the scaling factor $A$ in eq.~\ref{eq_nofP}. 
The values of these parameters  depend on the frequency and angular resolution of the survey.
With only three flux density bins, the model is degenerate and different combinations of parameters provide a comparable match to the data. With a scaling factor $A = 1$, a very low value of the median fractional polarization is required (Fig.~\ref{fig:counts}). More realistically, the scaling factor $A$ can be reduced to indicate that only a fraction of the radio sources detected in continuum have measurable polarization. Polarization measurements of a larger field or even deeper measurements in ELAIS-N1 would increase the number of bins of polarized sources, and make it possible to include several polarized source populations with different characteristics in the model. 

\section{Conclusions}\label{sec:conclusion}

In this work, we developed a new method to stack LOFAR datasets taken in different observing cycles with different frequency configurations. 
Stacking datasets from 19 epochs allowed us to reach a noise level in polarization of 19~$\mu$Jy~beam$^{-1}$ in the central part of the final image of 25~deg$^2$ of the ELAIS-N1 field. To our knowledge, this is is the most sensitive polarization dataset obtained at 150~MHz so far. 
This allowed us to detect 33 polarized components in 31 sources in ELAIS-N1 (two of the sources are double-lobed radio galaxies in which polarization was detected from both lobes) and probe the sub-mJy population of polarized sources at low frequencies.

The number density of polarized sources found is 1.24 per square degree, which is approximately twice as high as what was found in the first LOFAR polarization survey of ELAIS-N1 by \cite{HerreraRuiz2021}, and three times more than in the LoTSS-DR2 RM grid \citep{lotssdr2}. 
The catalog has two parts: one resulting from a search in the whole 25~deg$^2$ field, and another resulting from a  polarization search towards sources known to be polarized at 1.4~GHz from the work of \cite{grant2010} and the RM catalog from NVSS \citep{taylor2009}. 
For these sources with prior information on polarization the detection threshold was lowered to 6$\sigma_{\rm QU}$. 
The catalog of 25 sources detected above 8$\sigma_{\rm QU}$ was used to construct polarized source counts down to a polarized point-source flux density of 200~$\mu$Jy. 
The observed polarized counts were modeled using the polynomial fit to source counts in total-intensity obtained for three LOFAR Deep Fields by \cite{Mandal2021}, convolved with a log-normal probability distribution function of fractional polarizations; the parameters of the model were partly taken from the statistical properties of the sample and adjusted to match the observed data points. 

The methods presented here may be used in future polarization studies of LOFAR deep and ultra-Deep Fields and for polarization data taken in other radio frequency bands. 
Our work has shown that the polarization fraction is higher at 6$''$ resolution than at 20$''$, and that the combination of stacking with imaging at higher angular resolution leads to a higher number of detections of polarized sources. Future promising work includes searching for polarization in LOFAR data at even higher angular resolution: sub-arcsecond resolution with the international baselines.

Whereas the frequency range of LOFAR allows for precise RM determinations, Faraday depolarization is a limiting factor. 
The POlarisation Sky Survey of the Universe's Magnetism (POSSUM), carried out in a band centred at $\sim$890~MHz with the Australian Square Kilometre Array Pathfinder (e.g. \citealt{2021PASA...38...20A}, \citealt{Vanderwoude2024AJ....167..226V}), 
as well as future observations in Band~1 and Band~2 of SKA-Mid (350-1050~MHz; 950-1760~MHz) will provide a trade-off between depolarization and precision on the RM values and will be powerful tools to construct large RM grids \citep{heald2020}.   

In a companion paper (Piras et al.; Paper~II) we will characterize the detected extragalactic polarized sources in ELAIS-N1 in terms of their morphologies in radio continuum, their redshifts, linear sizes, rest-frame luminosities and environments, and present the RM grid derived from the RM values obtained with LOFAR.   

\begin{acknowledgements}
We thank the referee for helpful comments. 
LOFAR, the Low Frequency Array designed and constructed by ASTRON, has facilities in several countries, that are owned by various parties (each with their own funding sources), and that are collectively operated by the International LOFAR Telescope (ILT) foundation under a joint scientific policy. 
This work was done within the LOFAR Surveys and the LOFAR Magnetism Key Science Projects. 
This research has made use of the NASA/IPAC Extragalactic Database (NED) which is operated by the Jet Propulsion Laboratory, California Institute of Technology, under contract with the National Aeronautics and Space Administration.
This research has made use of the VizieR catalog access tool, CDS, Strasbourg, France. The original description of the VizieR service was published in A\&AS 143, 23.
We have also made use of the table analysis software {\tt topcat} \citep{topcat}.
This research made use of {\tt Astropy}, a community-developed core
Python package for astronomy \citep{astropy}, of {\tt Matplotlib} \citep{Hunter2007}, and of
{\tt APLpy} \citep{RobitailleBressert2012}, an open-source astronomical
plotting package for Python. 
The processing of LOFAR data was enabled by resources provided by the Swedish National Infrastructure for Computing (SNIC) at Chalmers Centre for Computational Science and Engineering (C3SE) partially funded by the Swedish Research Council through grant agreement no.~2018-05973. 
SP would like to thank R.~Beck, T.~Carozzi and M.~C.~Toribio for useful comments and discussions.
SdP gratefully acknowledges support from the European Research Council (ERC) Advanced Grant, 789410.
SPO acknowledges support from the Comunidad de Madrid Atracción de Talento program via grant 2022-T1/TIC-23797. 
VJ acknowledges support by the Croatian Science Foundation for a project IP-2018-01-2889 (LowFreqCRO). 
AB was supported by funding from the German Research Foundation DFG, within the Collaborative Research Center SFB1491 "Cosmic Interacting Matters - From Source to Signal".
PNB is grateful for support from the UK STFC via grant ST/V000594/1. IP acknowledges support from INAF under the Large Grant 2022 funding scheme (project "MeerKAT and LOFAR Team up: a Unique Radio Window on Galaxy/AGN co-Evolution”).
\end{acknowledgements}

\bibliographystyle{aa}
\bibliography{aanda.bib}

\begin{appendix}

\section{The RM spread function}
\begin{figure}[h]
\includegraphics[width=\linewidth]{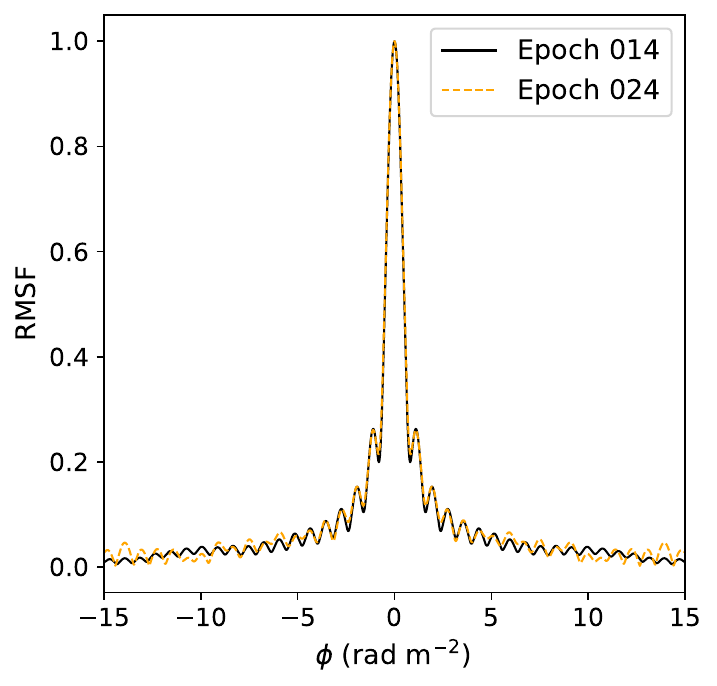}
\caption{Amplitude of the RM Spread Function (RMSF) corresponding to data taken in Cycle 2 (Epoch~014) in black line and Cycle 4 (Epoch~024) in dashed orange line.}
\label{fig:rmsf}
\end{figure}

\section{Catalog of polarized sources in the ELAIS-N1 field detected with LOFAR}

\begin{table*}
\caption{Catalog of polarized sources in the ELAIS-N1 field detected with LOFAR.}
\small
\label{table:det_cat}       
\begin{tabular}{l c c c c r r c c}  
\hline\hline  
Source ID & LOFAR ID & RA (J2000) & Dec (J2000) & $P_{\rm p}$ & RM & $I_{\rm p}$ & $\Pi$ & Pol. info\\
&   & (deg) &  (deg) &  (mJy beam$^{-1}$) & (rad m$^{-2}$) & (mJy beam$^{-1}$) & ($\%$) &\\  
(1) &(2) &(3) &(4) &(5) &(6) &(7) &(8) &(9)\\
\hline
\hline
01$^{(b,c)}$ & ILTJ155603.98+550056.8  & 239.0149 & 55.0158 &  1.02 & 1.68 $\pm$ 0.02 &  59.15 & 1.72 \\
02 & ILTJ155614.81+534814.8  & 239.0612 & 53.8012 &  0.54 & 9.71 $\pm$ 0.05 &  76.83 & 0.71 \\
03$^{(b,c)}$ & ILTJ155848.42+562514.4  & 239.7013 & 56.4209 &  1.19 & -5.83 $\pm$ 0.02 &  187.98 & 0.64 & 3\\
04$_{\rm A}^{(a,b,c)}$ & ILTJ160344.42+524228.0  & 240.9436 & 52.6961 &  2.69 & 19.68 $\pm$ 0.01 &  135.80 & 1.98 & 2\\
04$_{\rm B}^{(a,b,c)}$ & ILTJ160344.42+524228.0  & 240.9283 & 52.7150 &  9.07 & 21.78 $\pm$ 0.01 &  205.02 & 4.43 & 2  \\
05 & ILTJ160520.16+532837.2  & 241.3342 & 53.4771 &  0.33 & 19.12 $\pm$ 0.05 &  39.30 & 0.86  & 1 \\
06$^{(b,c)}$ & ILTJ160532.84+531257.4  & 241.3853 & 53.2159 &  0.66 & 18.43 $\pm$ 0.03 &  19.41 & 3.42 & 3\\
{\bf 07}$^{(a,b,c)}$ & {\bf ILTJ160538.33+543922.6}  & 241.4080 & 54.6551 &  6.23 & 6.062 $\pm$ 0.003 &  355.62 & 1.75 &  1,2,3\\
08$^{(c)}$ & ILTJ160603.11+534812.6  & 241.5271 & 53.8078 &  0.39 & 13.39 $\pm$ 0.03 &  48.38 & 0.82  & 1 \\
09$^{(b)}$ & ILTJ160725.85+553525.8  & 241.8578 & 55.5905 &  0.38 & -3.79 $\pm$ 0.03 &  16.87 & 2.27 \\
10 & ILTJ160820.72+561355.7  & 242.0866 & 56.2325 &  0.24 & -5.61 $\pm$ 0.05 &  54.61 & 0.45   & 1\\
11 & ILTJ160847.74+561119.0  & 242.1929 & 56.1893 &  0.28 & -6.24 $\pm$ 0.04 &  38.32 & 0.75   & 1\\
12$^{(a,b,c)}$ & ILTJ160908.39+535425.4  & 242.2803 & 53.9090 &  0.72 & 7.17 $\pm$ 0.02 &  18.68 & 3.86   & 1,3\\
13$_{\rm A}^{(c)}$ & ILTJ161112.81+543317.5  & 242.8025 & 54.5562 &  0.25 & 2.38 $\pm$ 0.04 &  3.77 & 6.62 \\
14$^{(b,c)}$ & ILTJ161240.15+533558.3  & 243.1678 & 53.5991 &  0.61 & 10.31 $\pm$ 0.02 &  33.66 & 1.82 & 3\\
15$^{(b)}$ & ILTJ161314.05+560810.8  & 243.2816 & 56.1327 &  0.32 & -4.80 $\pm$ 0.04 &  2.18 & 14.76   & 1,3\\
16 & ILTJ161529.67+545235.2  & 243.8747 & 54.8748 &  0.20 & -4.01 $\pm$ 0.05 &  3.85 & 5.17   & 1\\
17$^{(b,c)}$ & ILTJ161548.36+562030.1  & 243.9344 & 56.3492 &  0.53 & 1.95 $\pm$ 0.03 &  9.19 & 5.77 & 3\\
18$^{(b,c)}$ & ILTJ161623.79+552700.8  & 244.0991 & 55.4505 &  0.36 & -20.31 $\pm$ 0.03 &  9.00 & 4.07   & 1,3\\
19$^{(b)}$ & ILTJ161832.97+543146.3  & 244.6383 & 54.5344 &  0.30 & 2.91 $\pm$ 0.04 &  39.22 & 0.77 & 1,2\\
20$^{(a,b,c)}$ & ILTJ161919.70+553556.7  & 244.8332 & 55.6012 &  0.51 & -4.7 $\pm$ 0.02 &  33.55 & 1.52  & 1,3\\
21 & ILTJ162027.55+534208.8  & 245.1212 & 53.7011 &  0.43 & 3.29 $\pm$ 0.04 &  76.11 & 0.57 \\
22$^{(b)}$ & ILTJ162318.64+533847.4  & 245.8280 & 53.6447 &  0.54 & 2.31 $\pm$ 0.05 &  13.24 & 4.14 \\
23 & ILTJ162347.10+553207.2  & 245.9442 & 55.5346 &  0.36 & 5.91 $\pm$ 0.05 &  15.10 & 2.39   & 1\\
24$^{(a,b,c)}$ & ILTJ162432.20+565228.5  & 246.1343 & 56.8748 &  6.22 & 9.43 $\pm$ 0.01 &  138.47 & 4.50 & 2,3  \\
25$^{(a,b,c)}$ & ILTJ162634.18+544207.8  & 246.6426 & 54.7020 &  0.97 & 10.03 $\pm$ 0.02 &  71.80 & 1.35 & 1,2  \\
\hline
26 & ILTJ160936.45+552659.0  & 242.4032 & 55.4533 &  0.15 & -5.38 $\pm$ 0.07 &  10.17 & 1.55 & 1  \\
27 & ILTJ161037.49+532425.1  & 242.6549 & 53.4177 &  0.22 & 14.17 $\pm$ 0.07 &  3.18 & 7.02 & 1 \\
28$^{(b,c)}$ & ILTJ161057.72+553527.9  & 242.7404 & 55.5913 &  0.20 & -5.74 $\pm$ 0.05 &  36.72 & 0.56 & 1 \\
29$^{(c)}$ (13$_{\rm B}$)& ILTJ161120.73+543147.7  & 242.8390 & 54.5283 &  0.19 & 3.92 $\pm$ 0.05 &  8.62 & 2.19 & 1 \\
30 & ILTJ161340.99+524913.0  & 243.4168 & 52.8230 &  0.32 & 19.4 $\pm$ 0.07 &  29.41 & 1.11 & 1 \\
31$^{(b,c)}$ & ILTJ161537.86+534646.4  & 243.9072 & 53.7797 &  0.24 & 6.23 $\pm$ 0.06 &  18.50 & 1.32 & 1 \\
32 & ILTJ161859.41+545246.3  & 244.7448 & 54.8745 &  0.18 & -4.05 $\pm$ 0.06 &  23.05 & 0.82 & 1 \\
\hline                                
\end{tabular}
\begin{tablenotes}
\small
\item{
(1) Name of polarized source used in this paper; Source~7 (in bold) is the reference source;
Source 04$_{\rm A}$ and 04$_{\rm B}$ are two lobes of the same radio galaxy;
Source 13$_{\rm A}$ and 29 (or $13_{\rm B}$) are two lobes of the same radio galaxy.\\
(2) Name of associated total-intensity radio component in the \cite{Sabater2021} catalog;\\
(3), (4): Right ascension and declination of polarized source; \\
(5): Peak polarized intensity of the source in Cycle~2+Cycle~4 data; \\
(6): RM and RM error of polarized source in Cycle~2+Cycle~4 data; \\
(7): Stokes $I$ intensity in the image of \cite{Sabater2021} at the pixel of the peak polarized intensity;\\
(8): Degree of polarization in Cycle~2+Cycle~4 data; \\
(9): Flag to indicate whether the source is listed as polarized in other catalogs. 
      Value of 1 if the source is listed as polarized at 1.4 GHz in the DRAO ELAIS-N1 catalog \citep{grant2010}; 
      value of 2 if the source is listed in the NVSS RM catalog \citep{taylor2009};
      value of 3 if the source is listed as polarized at 150~MHz and 20$''$ resolution in the LOFAR work of \cite{HerreraRuiz2021}.\\
$^{(a)}$ Polarized source detected also in the reference epoch.\\
$^{(b)}$ Polarized source detected also in stacked Cycle~2.\\
$^{(c)}$ Polarized source detected also in stacked Cycle~4.\\
The bottom list (below the horizontal line) contains the polarized sources detected in the 6 to 8~$\sigma_{\rm QU}$ range.
}
\end{tablenotes}
\end{table*}

\end{appendix}

\end{document}